\documentclass[aps,pra,showpacs,superscriptaddress,amssymb,twocolumn]{revtex4-1}
\usepackage{graphicx}
\usepackage{appendix} 
\usepackage{hyperref}
\usepackage{bm,amsmath}
\usepackage{dcolumn}

\begin{document}

\rightline{\tt Physical Review A {\bf 91}, 062309 (2015)}

\vspace{0.2in}

\title{Universal quantum simulation with prethreshold superconducting qubits: Single-excitation subspace method}

\author{Michael R. Geller}
\email{mgeller@uga.edu}
\affiliation{Department of Physics and Astronomy, University of Georgia, Athens, Georgia 30602, USA}

\author{John M. Martinis}
\affiliation{Department of Physics and Google Inc., University of California, Santa Barbara, California 93106, USA}

\author{Andrew T. Sornborger}
\affiliation{Department of Mathematics, University of California, Davis, California 95616, USA}

\author{Phillip C. Stancil}
\affiliation{Department of Physics and Astronomy, University of Georgia, Athens, Georgia 30602, USA}
\affiliation{Center for Simulational Physics, University of Georgia, Athens, Georgia 30602, USA}

\author{Emily J. Pritchett}
\affiliation{HRL Laboratories LLC, 3011 Malibu Canyon Road, Malibu, California 90265, USA}

\author{Hao You}
\affiliation{Department of Physics and Astronomy, University of Georgia, Athens, Georgia 30602, USA}
\affiliation{Center for Simulational Physics, University of Georgia, Athens, Georgia 30602, USA}
\affiliation{Department of Chemistry and Physics, Georgia Regents University, Augusta, Georgia 30912, USA}

\author{Andrei Galiautdinov}
\affiliation{Department of Physics and Astronomy, University of Georgia, Athens, Georgia 30602, USA}

\date{\today}

\begin{abstract}
Current quantum computing architectures lack the size and fidelity required for universal fault-tolerant operation, limiting the practical  implementation of key quantum algorithms to all but the smallest problem sizes. In this work we propose an alternative method for general-purpose quantum computation that is ideally suited for such ``prethreshold" superconducting hardware. Computations are performed in the $n$-dimensional single-excitation subspace (SES) of a system of $n$ tunably coupled superconducting qubits. The approach is not scalable, but allows many operations in the unitary group SU($n$) to be implemented by a single application of the Hamiltonian, bypassing the need to decompose a desired unitary into elementary gates. This feature makes large, nontrivial quantum computations possible within the available coherence time. We show how to use a programmable SES chip to perform fast amplitude amplification and phase estimation, two versatile quantum subalgorithms. We also show that an SES processor is well suited for Hamiltonian simulation, specifically simulation of the Schr\"odinger equation with a real but otherwise arbitrary $n \! \times \! n$ Hamiltonian matrix. We discuss the utility and practicality of such a universal quantum simulator, and propose its application to the study of realistic atomic and molecular collisions.
\end{abstract}


\maketitle

\section{INTRODUCTION AND MOTIVATION}
\label{introduction section}

\subsection{The promise of quantum computation}

A universal quantum computer, if one could be built, would transform information technology by providing vastly increased computational power for certain specialized tasks, such as quantum simulation \cite{FeynmanIJTP82,LloydSci96,ZalkaProcRoySocLondA98,AspuruSci05,GeorgescuRMP14} and prime factorization \cite{ShorSIAMJC97,EkertRMP96}. Superconducting electrical circuits operating in the quantum regime \cite{SchoelkopfNat08,ClarkeNat08} have emerged as an extremely promising platform for realizing a large-scale, practical machine. Yet the quantum algorithms actually demonstrated to date---with any architecture---have been limited to only tiny, few-qubit instances \cite{ChuangPRL98,WeinsteinPRL01,VandersypenNat01,GuldeNat03,ChiaveriniNat04,PengPRA05,ChiaveriniSci05,NegrevergnePRA05,BrickmanPRA05,LuPRL07,LanyonPRL07,DiCarloNat09,LanyonNatChem10,BarreiroNat11,SchindlerSci11,LuPRL11,LanyonSci11,MariantoniSci11,ReedNat12,LuceroNatPhys12,MartinLopezNatPhoton12,FengSciRep13,BarendsNat14,ChowEtalNatComm14}. In superconducting circuit or circuit QED implementations, which benefit from the inherent scalability of modern solid-state electronics, this barrier in qubit number does not reflect any limitation of the underlying device fabrication or infrastructure requirements, but rather that larger problem sizes would also require longer computations and hence additional coherence. Quantum algorithms typically have (uncompiled) circuits that are spatially narrow but   temporally very deep. In this work we propose an alternative approach to superconducting quantum information processing that allows one to circumvent this restriction and realize much larger computations within the available coherence time.

A general-purpose quantum computer that is useful for practical applications must, of course, be error corrected and scalable. The standard model of an error-corrected quantum computer is the gate-based fault-tolerant universal quantum computer, where ``errors" acting on all device components and at any step during the computation can be corrected as long as they are weak enough---below an error threshold \cite{AharonovProcIEEE97,KitaevRussMathSurv97,KnillSci98}---and not highly correlated in space or time \cite{TerhalPRA05,AliferisQIC06,AharonovPRL06,NovaisPRA08}. Scalability means that the number of physical qubits required to perform a particular computation---the physical volume of the quantum computer---scales as a polynomial function (preferably linear) of the problem size. It also means that it is actually possible, in practice, to add more qubits. 

A realistic picture of an error-corrected superconducting quantum computer based on the surface code \cite{BravyiArxiv9811052,RaussendorfPRL07} is beginning to emerge \cite{FowlerPRA12}. The surface code is the most practical, best performing fault-tolerant approach known to date, and is especially amenable to implementaion with superconducting circuit technology. However, the resources required for a practical machine are considerable: Fowler {\it et al.}~[\onlinecite{FowlerPRA12}] estimated that factoring a 2000-bit number would require about $2 \! \times \! 10^8$ physical qubits, using Beauregard's modular exponentiation  \cite{BeauregardQIC03} and a surface code quantum computer operating at $99.9\%$ fidelity. If there was no decoherence or noise, and no errors of any kind to correct, then it would be possible to factor an $N$-bit number with the Beauregard algorithm using only $2N+3$ {\it ideal} qubits, or 4003 ideal qubits in the case considered. Thus, error correction imposes a physical qubit overhead of $2 \! \times \! 10^8 / 4003 \approx 5 \! \times \! 10^4$. Note that in quantifying the error-correction overhead here we distinguish between ideal (error free) qubits---the fictional entities usually appearing in quantum algorithms---and logical qubits, which must also include the many additional ancillas necessary for fault-tolerant gate implementation. Similarly, You {\it et al.} [\onlinecite{HaoGellerStancilPRA13}] estimated that it would take about $5 \! \times \! 10^6$ physical qubits to calculate the ground state energy of a 100-spin transverse-field Ising model to $99\%$ accuracy using the same $99.9\%$-fidelity surface code quantum computer. This well known statistical mechanics model maps especially well to a quantum computer, and for $N$ spins on a line would require only $N+1$ ideal qubits for a calculation of the ground state energy (using iterative phase estimation). So the physical/ideal ratio in this quantum simulation example is $5 \! \times \! 10^6 / 101 \approx 5 \! \times \! 10^4$, the same as for factoring. Therefore we expect that, in practice, surface code error correction will impose an overhead of
\begin{equation}
\frac{\rm \# \ physical \ qubits}{\rm \# \ ideal \ qubits} \approx 10^4,
\label{overhead estimate}
\end{equation}
where we have allowed for future optimization and other improvements.
Crudely, a factor of about 10 in the overhead estimate comes from replacing ideal qubits with enough logical qubits to both encode those ideal qubits and to distill the auxiliary states needed to perform fault-tolerant operations on them, and a factor of about $10^3$ comes from replacing each logical qubit with enough physical qubits to enable a sufficiently long computation.

\subsection{Prethreshold quantum computation}
\label{prethreshold section}

The complexity of building even a small fault-tolerant universal quantum computer suggests that this objective may take some time to achieve. In the meantime, experimental quantum information processing is limited to either the very small problem sizes discussed above, or to nonuniversal approaches such as analog quantum simulation \cite{GeorgescuRMP14,LewensteinAdvPhys07,BlochRMP08,HouckNatPhys12}, quantum annealing \cite{JohnsonNat11,BoixoNatPhys14}, or other special-purpose methods \cite{SornborgerSciRep12,AaronsonTheorComp13}.
In this work we label any quantum computation without an error-corrected universal quantum computer as {\it prethreshold}, referring to the threhold theorems of fault-tolerant quantum computation, because exceeding a fidelity threshold is a necessary condition for large-scale error correction.

Table \ref{prethreshold comparison table} compares three broad approaches to quantum computation with prethreshold hardware: {\it Small system} refers to gate-based computations with a few qubits, which have been used to test fundamental concepts of quantum information processing, demonstrate hardware functionality, and assess qubit and gate performance. The SES method is also general purpose, but should enable quantum speedup (this is discussed below). However neither approach is scalable. Analog quantum simulation and other scalable, special-purpose approaches trade universality for a faster route to speedup. 

The SES quantum computer described in this work is universal in the sense that it can implement any gate-based algorithm or quantum circuit. As a simulator it can directly emulate any (real) Hamiltonian, including time-dependent Hamiltonians. When we refer to a simulated Hamiltonian in this context we mean a Hamiltonian written in some basis---a real, symmetric {\it matrix} with no special structure. We assume that the Hamiltonian matrix to be simulated has been specified externally, as is typically the case when using a classical computer. The SES processor solves the Schr\"odinger equation defined by this Hamiltonian.

\begin{table}[htb]
\centering
\caption{Three approaches to prethreshold quantum computation and simulation. The left column lists the attributes achievable by an error-corrected universal quantum computer.}
\begin{tabular}{|c||c|c|c|}
\hline
 & small system &  SES method & analog QS/spec purp \\
\hline
{\sl scalable} & $\times$ & $\times$ & $\surd$ \\
\hline 
{\sl universal} & $\surd$ & $\surd$ & $\times$  \\
\hline 
{\sl speedup} & $\times$ & $\surd$ & $\surd$  \\
\hline 
{\sl arb accuracy} & $\times$& $\times$ & $\times$  \\
\hline 
{\sl arb runtime} & $\times$ & $\times$ & $\times$  \\
\hline 
\end{tabular}
\label{prethreshold comparison table}
\end{table}

The superconducting SES method introduced here has features in common with the single-photon protocols of Reck {\it et al.}~\cite{ReckPRL94} and Cerf {\it et al.}~\cite{CerfPRA98}, as both use only one excitation, and therefore do not utilize genuine entanglement. The optical realization uses a recursive algorithm to first decompose a given $n \! \times \! n$ unitary $U$ of interest into a sequence of SU(2) beam-splitter transformations. This decomposition determines an arrangement of beam splitters, phase shifters, and mirrors, that will unitarily transform $n$ input ports (optical modes) to $n$ output ports according to the desired $U$. However, the superconducting realization is better suited for quantum simulation than the optical approach because the Hamiltonian is directly programmed. In particular, to optically simulate Schr\"odinger evolution under a given Hamiltonian matrix $H$, one would have to first compute the evolution operator $e^{-iHt}$ on a classical computer, and then decompose it into beam-splitter transformations, but avoiding the classical computation of $e^{-iHt}$ (or the time-ordered exponential if $H$ is time dependent) is the motivation for the quantum computation in the first place \cite{opticalSesNote}. Neither the superconducting SES nor the single-photon optical approaches are scalable---they both require exponential physical resources---and should not be considered as viable alternatives to the standard paradigm of error-corrected universal quantum computation. But they are both suitable prethreshold methods. 

\section{QUANTUM COMPUTATION IN THE SES}

\subsection{Hardware model: The programmable SES chip}

Consider the following model of an array of $n$ coupled superconducting qubits,
\begin{eqnarray}
H_{\rm qc} = \sum_{i} \epsilon_i c_i^\dagger c_i
+ \frac{1}{2} \sum_{i i'} g_{ii'} \, \sigma^x_i\otimes\sigma^x_{i'},
\label{QC model}
\end{eqnarray}
written in the basis of uncoupled-qubit eigenstates. Here $ i,i' = 1, 2, \dots, n,$ and
\begin{equation}
c \equiv 
\begin{pmatrix}
0 & 1 \\
0 & 0 \\
\end{pmatrix}.
\label{c definition}
\end{equation}
The $\epsilon_i$ are qubit transition energies and the $g_{ii'}$ are qubit-qubit interaction strengths; both are assumed to be tunable. (Factors of $\hbar$ are suppressed throughout this paper.) $g_{ii'}$ is a real, symmetric matrix with vanishing diagonal elements. We also require microwave pulse control of at least one qubit, and simultaneous readout (projective measurement in the diagonal basis) of every qubit. The model (\ref{QC model}) describes a fully connected network or complete graph of qubits, which we refer to as an {\it SES processor}. This should be contrasted with {\it local} quantum computer models that have coupling only between nearby qubits (nearest neighbors, for example). The SES method can be applied with a wide variety of qubit-qubit interaction types (see Appendix~\ref{general coupling types section}), but without loss of generality we restrict ourselves here to the simple $\sigma^x \otimes \sigma^x$ coupling of (\ref{QC model}). Alternatively, tunably coupled resonators (with tunable frequencies) can be used instead of qubits \cite{resonatorNote}. Although we assume an architecture based on superconducting circuits (or circuit QED), the SES method might apply to other future architectures as well.

The quantum computer model (\ref{QC model}) might be considered unscalable, because of the $O(n^2)$ tunable coupling circuits and wires, a position that we also adopt here. In gate-based universal quantum computation, the fully connected and local quantum computer models are equivalent in the sense that any quantum circuit implemented by a fully connected quantum computer can be implemented by a local quantum computer after adding chains of {\sf SWAP} gates, which only introduce polynomial overhead. However, this equivalence is restricted to the standard gate-based approach and does not apply here.

Superconducting qubits have been reviewed in Refs.~\cite{SchoelkopfNat08} and \cite{ClarkeNat08}. Although the model (\ref{QC model}) can be realized with several qubit designs, the transmon qubit \cite{KochPRA07} currently has the best performance \cite{PaikPRL11,RigettiPRB12,BarendsPRL13,ChenEtalPRL14}. For concreteness we assume a qubit frequency $\epsilon/2\pi$ in the range of $5.45$ to $5.55 \, {\rm GHz}$ and coupling strength $g/2\pi$ in the range $-50$ to $50 \, {\rm MHz}$. An $n$-qubit SES processor also requires $n(n-1)/2$ coupler circuits and the associated wires or waveguides. A variety of tunable couplers can be used for this purpose. Here we consider a modification of the tunable inductive coupler developed by Chen {\it et al.}~\cite{ChenEtalPRL14} for superconducting Xmon qubits; this design has been demonstrated to implement tunability without compromising high coherence. Our modification replaces the direct electrical connection of each qubit to a coupler circuit wire with an inductive coupling to the wire, which scales better. An SES chip layout that avoids excessive crossovers is illustrated in Fig.~\ref{SES coupler circuit figure}. The tunable interaction strength $g$ for this coupler design is derived in Appendix \ref{tunable coupler circuit section}.

\begin{widetext}

\begin{figure}
\includegraphics[width=15.0cm]{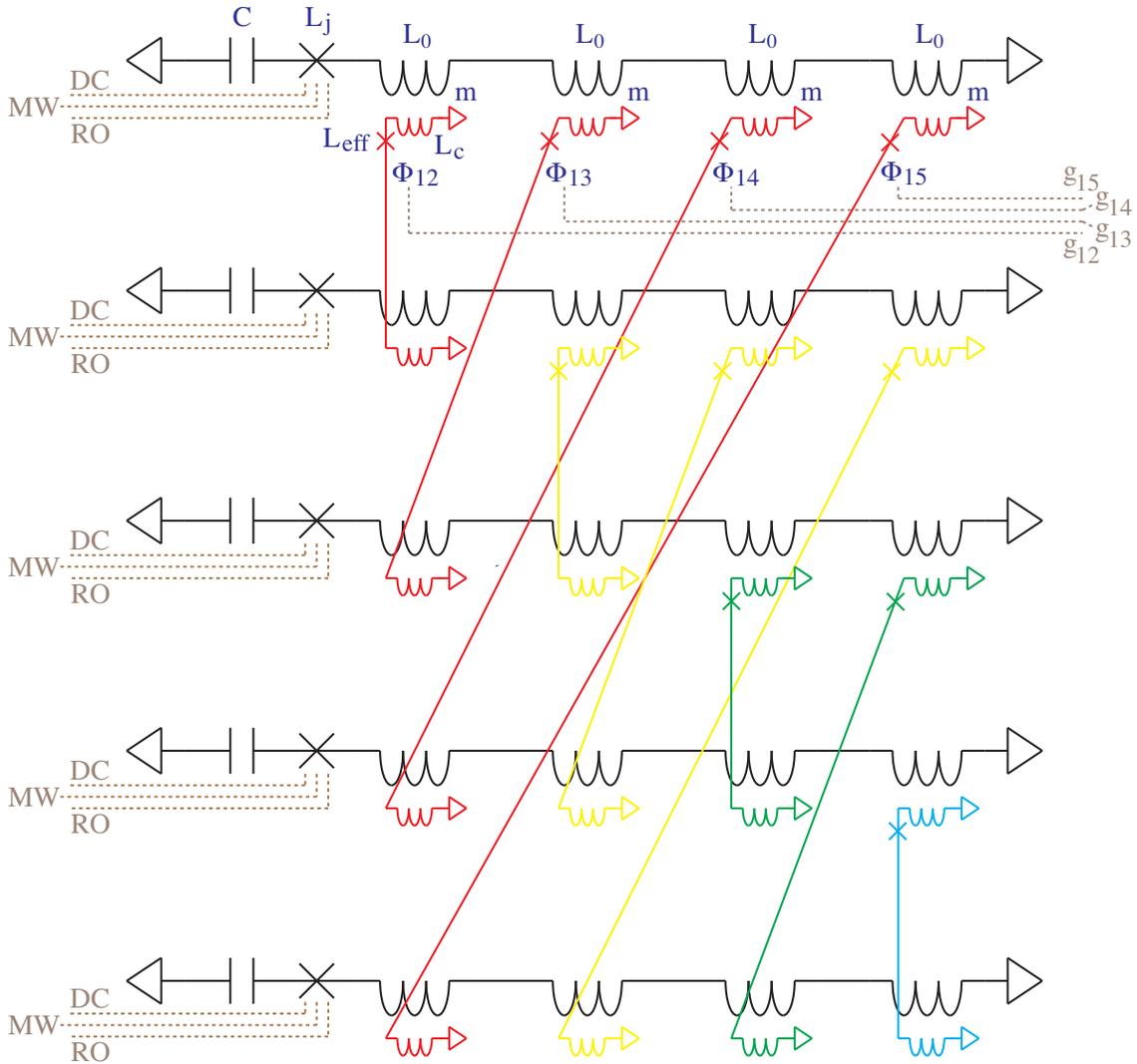} 
\caption{(Color online) Circuit layout for $n \! = \! 5$ SES processor, with the crosses representing Josephson junctions. Each horizontal circuit is an Xmon qubit with capacitance $C$, tunable junction inductance $L_{\rm j}$, and $n-1$ additional coils (each with self-inductance $L_0$ and mutual inductance $m$) for coupling to other qubits. Dotted lines indicate dc and microwave control lines for each qubit, as well as readout circuits. Each coupler wire contains a Josephson junction with inductance $L_{\rm c}$ tuned by a magnetic flux $\Phi$. Control lines for SES matrix elements $g_{12}, \dots , g_{15}$ are also indicated. This circuit is discussed further in Appendix \ref{tunable coupler circuit section}.}
\label{SES coupler circuit figure}
\end{figure} 

\end{widetext}

\subsection{Single-excitation subspace}

The idea we explore in this paper is to perform a quantum computation in the $n$-dimensional single-excitation subspace of the full $2^n$-dimensional Hilbert space. This is the subspace spanned by the computational basis states 
\begin{equation}
\big| i \big) \equiv c_i^\dagger \big|00 \cdots 0 \big\rangle = \big|0 \cdots 1_i  \cdots 0\big\rangle, 
\label{SES states}
\end{equation}
with $i=1,2,\dots,n.$ We call the set of $|i)$ the SES {\it basis} states. It is simple to prepare the quantum computer in an SES basis state from the ground state $|00\cdots0\rangle$, and it will remain there with high probability if the following conditions are satisfied:
\begin{enumerate}

\item The coupling strengths $|g_{ii'}|$ are much smaller than the $\epsilon_i$, which is usually well satisfied in superconducting circuits. 

\item Single-qubit operations such as $\pi$ and $\pi/2$ rotations about the $x$ or $y$ axes are not used during the computation. However, $2 \pi$ rotations are permitted and are very useful (these can be implemented as $z$ rotations, which do not require microwaves). $\pi$ rotations about $x$ or $y$ can be used to prepare SES basis states from the system ground state $|00 \cdots 0\rangle$.

\item The quantum computation time is significantly shorter than the single-qubit population relaxation time $T_1$.

\end{enumerate}

An SES pure state is of the form
\begin{equation}
\big| \psi \big\rangle = \sum_{i=1}^n a_i \, \big| i \big) , \  \ \ \  \ \ \sum_{i=1}^n|a_i |^2 = 1.
\label{general SES state}
\end{equation}
For example, the states (\ref{general SES state}) include the maximally entangled $W\!$-type state
\begin{eqnarray}
\big|{\rm unif} \big\rangle &\equiv&
\frac{| 1 ) +  | 2) + \cdots + | n)}{\sqrt{n}} \nonumber \\
&=& \frac{|10\cdots0\rangle +  |01\cdots0\rangle+ \cdots + |00\cdots1\rangle}{\sqrt{n}}. 
\label{unif state}
\end{eqnarray}
Although the state (\ref{unif state}) is entangled, and could be used to violate Bell's inequality, the entanglement is somewhat artificial \cite{LLoydPRA99} as there is only one ``particle".

\subsection{SES Hamiltonian}
\label{SES Hamiltonian section}

The advantage of working in the SES can be understood from the following expression for the SES matrix elements of model (\ref{QC model}), namely
\begin{equation}
{\cal H}_{ii'}  \equiv \big( i \big| H_{\rm qc} \big| i' \big)  = \epsilon_i \,  \delta_{ii'}  + g_{ii'}.
\label{SES hamiltonian}
\end{equation}
Because the diagonal and off-diagonal elements are directly and independently controlled by the qubit frequencies and coupling strengths, respectively, we have a high degree of programmability of the SES component of the quantum computer's Hamiltonian. This property allows many $n$-dimensional unitary operations to be carried out in a single step, bypassing the need to decompose into elementary gates, and also enables the direct quantum simulation of real but otherwise arbitrary time-dependent Hamiltonians. However, we have some restrictions:
   
\begin{enumerate}

\item ${\cal H}_{ii'}$ is always real, whereas the most general Hamiltonian matrix is complex Hermitian. The experimentally available control parameters, consisting of $n$ qubit frequencies and $n(n-1)/2$ coupling strengths, are sufficient to control the $n(n+1)/2$ independent parameters of an $n \! \times \! n$ real symmetric matrix.

\item There are experimental limitations on the range of values that the $\epsilon_i$ and $g_{ii'}$ can take. We define $g_{\rm max}$ to be the magnitude of the largest coupling available in a particular experimental realization; a current realistic value is about $50 \, {\rm MHz}$. 

\end{enumerate}
We will leave the discussion of possible generalizations to complex SES Hamiltonians for future work. The limitations on the ranges of the $\epsilon_i$ and $g_{ii'}$ do not, by themselves, limit the class of Hamiltonians that can be  simulated, because a model Hamiltonian intended for simulation is first rescaled to conform to that of the SES chip (this is explained below).

It will be useful to refer to a ``typical" SES Hamiltonian ${\cal H}$, which we assume to have the following properties: ${\cal H}$ is a real, symmetric matrix with each element taking values in the range $-g_{\rm max}$ to $g_{\rm max}$. This form follows from 
(\ref{SES hamiltonian}) after removing an unimportant term proportional to the identity matrix,
\begin{equation}
{\cal H} \rightarrow  {\cal H} - \omega_{\rm ref}  I,
\end{equation}
where $\omega_{\rm ref}$ is a convenient (possibly time-dependent) reference frequency. Then we assume that the qubit frequencies $\epsilon_i$ can be tuned within $\pm g_{\rm max}$ of $\omega_{\rm ref}$, and we assume that the couplers can be tuned between $-g_{\rm max}$ to $g_{\rm max}$. A possible choice for $\omega_{\rm ref} $ is the mean qubit frequency $(1/n) \sum_i  \epsilon_i$. 

\begin{figure}
\includegraphics[width=8.0cm]{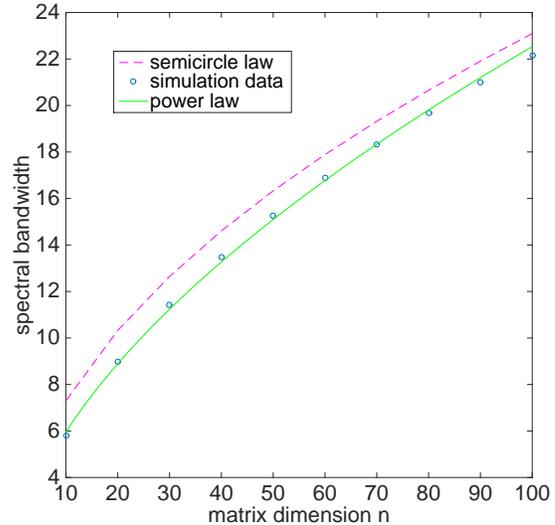} 
\caption{(color online) Spectral bandwidth of $K$ matrices versus $n$. Data (open circles) are averaged over 1000 random instances of $K$. The solid line is the function $1.58 \! \times \! n^{0.58}$. The dashed line is the function $(4/\sqrt{3}) \! \times \! \! \sqrt{n}.$}
\label{bandwidth100 figure}
\end{figure} 

It will also be useful to consider the statistical properties of an ensemble of typical SES matrices. We can always write a time-independent SES Hamiltonian in the {\it standard form}
\begin{equation}
{\cal H} = g_{\rm max} \, K,
\label{SES matrix decomposition}
\end{equation}
where $K$ is real symmetric matrix with every element satisfying
\begin{equation}
-1 \le K_{ii'} \le 1.
\label{K matrix condition}
\end{equation}
We define a real random matrix ensemble of dimension $n$ as follows: The $n$ diagonal elements are independent random variables $K_{ii}$, each uniformily distributed between $-1$ and $1$. The $n(n-1)/2$ elements $K_{i<i'}$ are independent random variables also uniformily distributed between $-1$ and $1$. The remaining elements $K_{i>i'}$ are fixed by the symmetry requirement. The standard deviation of each element is $\sigma_{\scriptscriptstyle \! K} = 1/\sqrt{3}.$

\begin{figure}
\includegraphics[width=8.0cm]{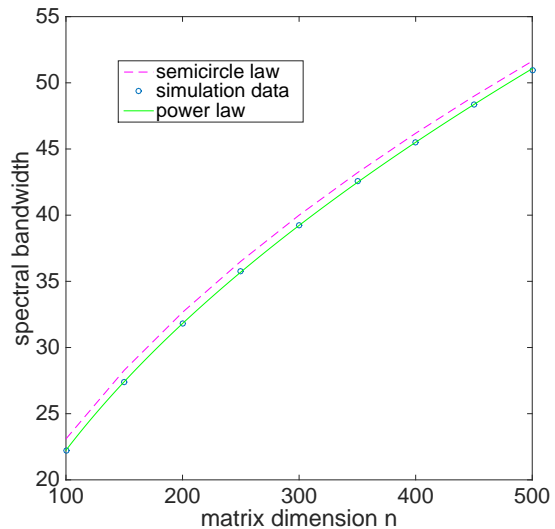} 
\caption{(color online) Spectral bandwidth for larger $n$. The solid line is the function $2.06 \! \times \! n^{0.52}$. The dashed line is the function $(4/\sqrt{3}) \! \times \! \! \sqrt{n}.$}
\label{bandwidth500 figure}
\end{figure} 

The first property we study is the mean spectral bandwidth of $K$, the difference between the largest and smallest eigenvalues. Let $\lambda_1 \le \lambda_2 \le \cdots \le \lambda_n$ be the ordered eigenvalues of $K$. From the Wigner semicircle law  \cite{MehtaRandomMatrices} we expect that, in the large $n$ limit,
\begin{equation}
\overline{
\lambda_{n} - \lambda_{1} } \rightarrow 
4 \sigma_{\scriptscriptstyle \! K} \sqrt{n} = \textstyle{\frac{4}{\sqrt{3}}} \, n^\frac{1}{2},
\label{K bandwidth formula Wigner}
\end{equation}
where the overbar denotes averaging over the ensemble defined above. The bandwidth of typical SES states in an $n$-qubit  processor therefore scales at large $n$ as
\begin{equation}
\overline{
E_{\rm max} - E_{\rm min} } \rightarrow \textstyle{\frac{4}{\sqrt{3}}} \, g_{\rm max} \, n^\frac{1}{2}.
\label{bandwidth formula Wigner}
\end{equation}
In Figs.~\ref{bandwidth100 figure} and \ref{bandwidth500 figure} we plot the simulated bandwidth of $K$ as a function of $n$, and compare the simulation data with the asymptotic form (\ref{K bandwidth formula Wigner}). From Fig.~\ref{bandwidth100 figure} we conclude that for modest SES matrix sizes,
\begin{equation}
\overline{
E_{\rm max} - E_{\rm min} } \approx 1.58 \, g_{\rm max} \,n^{0.58}.
\label{bandwidth formula emperical}
\end{equation}

\begin{figure}
\includegraphics[width=8.0cm]{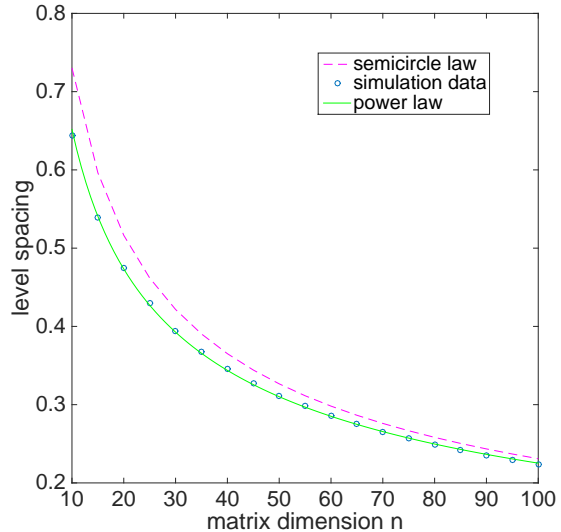} 
\caption{(color online) Level spacing of $K$ matrices versus $n$. Data (open circles) are averaged over 1000 random instances of $K$. The solid line is the function $1.89 \! \times \! n^{-0.46}$. The dashed line is the function $4 / \! \sqrt{3n}.$}
\label{levelspacing100 figure}
\end{figure}

The second property we study is the mean level spacing of $K$. Let 
\begin{equation}
\Delta \lambda \equiv \frac{1}{n-1} \sum_{i=1}^{n-1} \lambda_{i+1} - \lambda_{i}
\label{level spacing definition}
\end{equation}
be the mean spacing between adjacent eigenvalues. Averaging (\ref{level spacing definition}) over the ensemble defined above, we expect that in the large $n$ limit
\begin{equation}
{\overline{\Delta \lambda}} \approx \frac{ \overline{
\lambda_{n} - \lambda_{1} } }{n} \rightarrow
\textstyle{\frac{4}{\sqrt{3}}} \, n^{-\frac{1}{2}}.
\label{mean level spacing formula asymptotic}
\end{equation}
In Fig.~\ref{levelspacing100 figure} we plot the simulated average level spacing of $K$ as a function of $n$, and compare the simulation data with the asymptotic form (\ref{mean level spacing formula asymptotic}). From Fig.~\ref{levelspacing100 figure} we conclude that for modest SES matrix sizes,
\begin{equation}
\overline{\Delta E} \approx 1.89 \, g_{\rm max} \,n^{-0.46}.
\label{levelspacing formula emperical}
\end{equation}
The results (\ref{bandwidth formula emperical}) and (\ref{levelspacing formula emperical}) give two relevant energy scales present in a typical SES spectrum.

Any unitary quantum circuit or operation acting on $q$ qubits can be mapped to and implemented on an SES chip with $n\!=\!2^q$ qubits (this exponential growth of $n$ is what makes the SES method unscalable). We can say that the SES processor {\it simulates} the $q$-qubit system, with the advantage of being able to perform multi-qubit operations in a single step. This feature provides the computational advantage of the SES approach and is illustrated throughout this paper. It will be useful to specify an explicit one-to-one mapping between the bases of the associated Hilbert spaces, which we take to be
\begin{eqnarray}
|0 0 \cdots 0\rangle &\longleftrightarrow& |1) = |10 \cdots 0\rangle, \nonumber \\
|0 0 \cdots 1\rangle &\longleftrightarrow& |2)=|01 \cdots 0\rangle, \nonumber \\
&\vdots& \nonumber \\
\underbrace{|1 1 \cdots 1 \rangle}_{q \, {\rm qubits}} &\longleftrightarrow& |2^q) = \underbrace{|00 \cdots 1\rangle}_{n=2^q \, {\rm qubits}}. 
\label{natural mapping definition}
\end{eqnarray}
The left-hand-sides are the standard computational basis states of the simulated $q$-qubit system (not to be confused with the computational basis states of the $n$-qubit SES processor). Similarly, any unitary quantum circuit or operation acting on $q$ $d$-level qudits can be mapped to and implemented on an SES processor with $n\!=\!d^q$ qubits; the natural mapping is a straightforward generalization of (\ref{natural mapping definition}).

The operation of a real SES chip will be nonideal, and it is important to consider the effects of decoherence and other errors on its performance. This is discussed in detail in Sec.~\ref{error section}. The main conclusion is that although decoherence and unitary control errors do limit the accuracy of an SES computation or simulation, the effects of decoherence are much less restrictive here than with the standard gate-based approach (hence the ability to implement larger problem sizes). In practice, the complexity of fabricating a large programmable SES chip will likely limit its application before decoherence does.

\section{APPLICATIONS OF THE SES METHOD}

\subsection{Uniform state preparation}

\begin{figure}
\includegraphics[width=5.5cm]{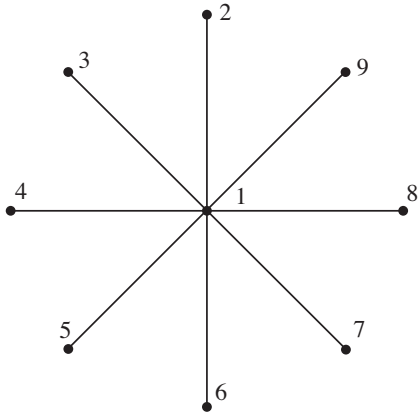} 
\caption{Star network for $n=9$ uniform state preparation.}
\label{star network figure}
\end{figure}

Our first example will be to generate the entangled state (\ref{unif state}) in a single step: Consider the real
$n \! \times \! n$ Hamiltonian
\begin{equation}
{\cal H} = g_{\rm max} K_{\rm star},
\label{uniform state prep H}
\end{equation}
where
\begin{equation}
K_{\rm star} \equiv
\begin{pmatrix}
1 & \frac{1}{2} & \frac{1}{2} & \cdots & \frac{1}{2} \\
\frac{1}{2} & 0 & 0 & \cdots & 0 \\
\frac{1}{2} & 0 & 0 & \cdots & 0 \\
\vdots & \vdots & \vdots  & \ddots & \vdots \\
 \frac{1}{2} & 0 & 0 & \cdots & 0 \\
\end{pmatrix}.
\label{K star definition}
\end{equation}
The Hamiltonian (\ref{uniform state prep H}) describes a graph where qubit 1 is symmetrically coupled to all other qubits, which are themselves uncoupled (a star network). The case of $9$ qubits is shown in Fig.~\ref{star network figure}. 

The SES chip is initially prepared in basis state $|1)$. Only two eigenfunctions---let's call them $|\psi_\pm\rangle$---have overlap with $|1)$, so the evolution is effectively a two-channel problem. The spectrum is as follows: States $|\psi_\pm\rangle$ have energy $E_\pm = g_{\rm max}(1\pm \sqrt{n})/2 $; all other eigenfunctions are degenerate with $E\!=\!0$. Evolution for half a period corresponding to the splitting $\sqrt{n} \, g_{\rm max}$, namely
\begin{equation}
t_{\rm qc} = \frac{\pi}{\sqrt{n} \, g_{\rm max}},
\label{t unif state prep}
\end{equation}
leads to the desired operation
\begin{equation}
e^{- i {\cal H} t_{\rm qc}} \big| 1 \big)  = \exp \! \! \left[ \! -i \frac{\pi}{\sqrt{n}}
\begin{pmatrix}
1 & \frac{1}{2}  & \cdots & \frac{1}{2} \\
\frac{1}{2}  & 0 & \cdots & 0 \\
\vdots & \vdots  & \ddots & \vdots \\
 \frac{1}{2} &  0 & \cdots & 0 \\
\end{pmatrix}
\! \right] \! \! \big| 1 \big) = \big|{\rm unif}\big\rangle,
\label{uniform state preparation identity}
\end{equation}
apart from a phase. This can be implemented in a few ns with superconducting circuits. 

We would like to make a few remarks that apply to this application, as well as to many others: First, the magnitude of the interaction strength used in (\ref{uniform state prep H}) is arbitrary; any convenient interaction strength $g_0$ satisfying $0 < g_0 \le g_{\rm max}$ is sufficient. To make the operation as fast as possible, however, we have choosen $g_0\!=\!g_{\rm max}$. Second, it is not necessary to use a time-independent interaction strength. Any single-step ``pulse sequence" of the form
\begin{equation}
e^{- i {\cal H} t_{\rm qc}},
\end{equation}
with
\begin{equation}
{\cal H}=g_0 K
\end{equation}
and $K$ a constant (time-independent) matrix, satisfies an {\it area theorem} 
\begin{equation}
T \, e^{- i \int \! g(t) K \, dt} 
= e^{- i {\cal H} t_{\rm qc}} ,
\label{area theorem identity}
\end{equation}
where
\begin{equation}
\int \! g(t) \, dt  =  g_{0} \, t_{\rm qc}
\label{area theorem integral}
\end{equation}
and $T$ is the time-ordering operator. The identity (\ref{area theorem identity}) implies that any time-dependent coupling $g(t)$ satisfying (\ref{area theorem integral}) can be used, simplifying experimental implementation.

\subsection{Grover search algorithm}

Next we show how to use a programmable SES chip to implement the Grover search algorithm \cite{GroverPRL97}, which introduced the powerful amplitude amplification technique that has led to speedup for many other algorithms.
Grover's procedure for a single marked state $|i)$ in a database of size $n$ is
\begin{equation}
\big( W O_i \big)^{\beta} |{\rm unif}\rangle \approx |i), \ \ {\rm with} \ \  
\beta = \big\lfloor {\textstyle \frac{\pi}{4}} \sqrt{n} \big\rfloor.
\label{grover identity}
\end{equation}
Here 
\begin{eqnarray}
W &\equiv& 2 \big|{\rm unif} \big\rangle \big\langle{\rm unif}\big| - I \nonumber \\
&=& \frac{1}{n}
\begin{pmatrix}
2-n & 2 & 2 & \cdots & 2 \\
2 & 2-n & 2 & \cdots & 2 \\
2 &  2 & 2-n & \cdots & 2 \\
\vdots & \vdots &\vdots & \ddots & \vdots \\
2 & 2 & 2 & \cdots & 2-n \\
\end{pmatrix},
\label{inversion operator}
\end{eqnarray}
is a unitary operator that performs an inversion about the average,
\begin{equation}
O_i  \equiv
\begin{pmatrix}
1 &  &  &  &   & \\
 & 1 &  &  &  & \\
 &  & \ddots &  & & \\
 &  & &  -1 & & \\
 &  & &   & \ddots & \\
 &  & &   & & 1 \\
\end{pmatrix},
\label{oracle operator}
\end{equation}
is the oracle, a diagonal matrix with the $i$th element equal to $-1$ and the others equal to $1$, and  $|{\rm unif})$ is the uniform superposition (\ref{unif state}). 

The $W$ operator (\ref{inversion operator}) can be implemented in a single step by using the SES Hamiltonian ${\cal H} = g_{\rm max} K_{\rm full},$ with
\begin{equation}
K_{\rm full} \equiv
\begin{pmatrix}
0 & 1 & 1 & \cdots & 1 \\
1 & 0 & 1 & \cdots & 1 \\
1 & 1 & 0 & \cdots & 1 \\
\vdots  &  \vdots & \vdots  & &  \vdots  \\
1& 1 & 1 & \cdots & 0 \\
\end{pmatrix},
\label{Kfull definition}
\end{equation}
for a time
\begin{equation}
t_{\rm qc} = \frac{\pi}{n \, g_{\rm max}}.
\label{t inversion operator}
\end{equation}
This leads to the desired operation
\begin{equation}
\exp \! \left[ -i \, \frac{\pi}{n}
\begin{pmatrix}
0 & 1  & \cdots & 1 \\
1 & 0  & \cdots & 1 \\
\vdots &  \vdots  & \ddots & \vdots  \\
1& 1  & \cdots & 0 \\
\end{pmatrix}
\right] = W,
\end{equation}
up to a phase factor. 

The oracle (\ref{oracle operator}) can be simply generated by a $2\pi$ rotation on qubit $i$. This $2\pi$ rotation can be implemented as a $z$ rotation, which does not require microwaves. Each iteration of the amplitude amplification can therefore be implemented in just two steps, for any $n$, allowing even small SES chips to perform  computations that would otherwise require thousands of elementary gates.

\subsection{Eigenvalue estimation}

Next we show how to use a programmable SES processor to implement energy eigenvalue estimation, an application of the important phase estimation algorithm \cite{KitaevArxiv9511026,CleveProcRoySocLondA98,AbramsPRL99} that is used in many other applications. This example also illustrates how to translate an algorithm expressed in quantum circuit language to an SES protocol. 

\begin{figure}
\includegraphics[width=7.0cm]{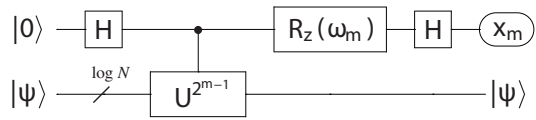} 
\caption{Quantum circuit to compute the $m$th bit of $\phi$. Here {\sf H} is the Hadamard gate, $U=e^{-iHt}$ with $H$ the model Hamiltonian and $t$ the evolution time, and $R_z$ is a $z$-rotation. $|\psi\rangle$ is an eigenfunction of $H$. The last operation is measurement of the first qubit in the diagonal basis; the result is $x_m \in \{0,1\}$.}
\label{IPE circuit figure}
\end{figure}

The eigenvalue estimation procedure calculates an $M$-bit estimate of the phase $\phi$ of the eigenvalue $e^{\displaystyle -i2\pi\phi}$ accumulated by an eigenfunction $|\psi\rangle$ under the action of $e^{- i H t} \! . \,$ If the evolution time $t$ is chosen to satisy $t < 2 \pi /E,$ the eigenvalue $E$ (assumed to be positive) can be calculated from $E = 2\pi\phi/t$. To reduce the number of required qubits we use the iterative phase estimation circuit \cite{DobsicekPRA07} shown in Fig.~\ref{IPE circuit figure}, which uses only a single ancilla. As the number $M$ of desired bits of precision increases, one either performs a longer quantum computation---reusing the eigenfunction $|\psi\rangle$---or performs $M$ computations in series, each requiring an eigenfunction preparation step. 

The algorithm measures $M$ bits of $\phi$ one at a time, beginning with the least significant bit $x_M$, and working backwards to the most significant bit $x_1$. Each step (except for the first) uses knowledge of the previously measured bits. We denote the bit being measured in a given step by $m$, with $m = M, M-1 , M-2 , \, \cdots , 1.$ The circuit for step $m$ is shown in Fig.~\ref{IPE circuit figure}, where the rotation angle is
\begin{equation}
\omega_m = \pi \sum_{j = m+1}^M \frac{x_j}{2^{j-m}},
\label{omega formula}
\end{equation}
which depends on the values of the previously measured bits $x_{m+1}, \cdots, x_M$.

The main practical difficulty with prethreshold applications of phase estimation is implementation of the
\begin{equation}
{\rm controlled} \! - \! e^{-iH 2^{m-1} t}
\label{controlled evolution operator}
\end{equation}
operation, which typically requires a Trotter approximation (and, in addition, a sparse Hamiltonian). However, the SES method allows {\it any} controlled unitary 
\begin{equation}
{\rm controlled} \! - \! U
\label{general controlled unitary}
\end{equation}
to be implemented in a single step when $U$ can (which is possible when $U$ is symmetric). To see this, assume that $U=e^{-iA}$, where $A$ is a real $N \! \times \! N$ matrix, and write the $2N \! \times \! 2N$ matrix (\ref{general controlled unitary}) as
\begin{equation}
{\rm controlled} \! - \! U = |0\rangle \langle 0 | \otimes I_{N \! \times \! N} \, + \, |1\rangle \langle 1 | \otimes e^{-iA},
\label{general controlled unitary identity}
\end{equation}
where $I$ is the identity matrix and where we take the first qubit to be the control. Next map the $2N$-dimensional Hilbert space to the SES processor according to
\begin{eqnarray}
|0 \rangle \otimes |1\rangle  \ &\longleftrightarrow& \ |1), \nonumber \\
|0 \rangle \otimes |2\rangle \  &\longleftrightarrow& \ |2), \nonumber \\
&\vdots & \nonumber \\
|0 \rangle \otimes |N \rangle  &\longleftrightarrow& \ |N), \nonumber \\
|1 \rangle \otimes |1 \rangle \ &\longleftrightarrow& \ |N+1), \nonumber \\
|1 \rangle \otimes |2 \rangle \ &\longleftrightarrow& \ |N+2), \nonumber \\
&\vdots & \nonumber \\
|1 \rangle \otimes |N \rangle  &\longleftrightarrow& \ |2N).
\label{controlled U mapping definition}
\end{eqnarray}
The operation (\ref{general controlled unitary}) can therefore be written as
\begin{equation}
{\rm controlled} \! - \! e^{-iA} = 
\exp \! \left[ -i  \bigg(
\begin{array}{c|c}
{\bf 0} & {\bf 0} \\
\hline 
{\bf 0} & A
\end{array} 
\bigg) \right] \! ,
\label{controlled exp A identity}
\end{equation}
which can be implemented by an SES processor in a single step. The elements ${\bf 0}$ and $A$ on the right-hand-side of (\ref{controlled exp A identity}) are each $N \! \times \! N$ matrices, with ${\bf 0}$ the zero (null) matrix.

We turn now to the SES eigenvalue estimation protocol: Let $H$ be a real $N \! \times \! N$ {\it model} Hamiltonian on which we wish to perform phase estimation, and denote the basis of $H$ by $\lbrace |1\rangle, |2\rangle, \cdots , |N\rangle \rbrace$. The SES implementation requires $n=2N$ qubits. The first objective in the protocol is to prepare the initial state
\begin{equation}
|0\rangle \otimes |\psi\rangle
\label{phase estimation initial state}
\end{equation} 
of Fig.~\ref{IPE circuit figure}, where $|\psi\rangle$ is an eigenfunction of $H$. We will perform the state preparation adiabatically, which is restricted to states of minimum or maximum energy; here we prepare the ground state of $H$ and estimate the ground state energy $E$.

Adiabatic ground state preparation is usually implemented by programming a convenient initial Hamiltonian $H_0$ that does not commute with $H$, relaxing into the ground state of $H_0$, and then slowly changing the system Hamiltonian from $H_0$ to $H$. However, in the SES approach it is necessary to use {\it nonequilibrium} adiabatic evolution, because the physical ground state $|0\rangle^{\otimes n}$ is outside the SES. The processor is initially prepared in the basis state $|1)$. The next step is to produce the SES state equivalent to 
\begin{equation}
|0\rangle \otimes \frac{|1\rangle + |2\rangle + \cdots + |N\rangle}{\sqrt N},
\label{input state to adiabatic preparation}
\end{equation}
which, according to the map (\ref{controlled U mapping definition}), is
\begin{equation}
\frac{|1) + |2) + \cdots + |N)}{\sqrt N}.
\label{input SES state to adiabatic preparation}
\end{equation}
Note that (\ref{input SES state to adiabatic preparation}) is a uniform superposition of the first {\it half} of SES basis states. To prepare this we use a variation of (\ref{uniform state preparation identity}), namely
\begin{equation}
e^{-i {\cal H} t_{\rm qc}}  \big|1\big) =
\frac{|1) + |2) + \cdots + |N)}{\sqrt N},
\label{half uniform state preparation identity}
\end{equation}
where
\begin{equation}
{\cal H} = g_{\rm max} \left( \begin{array}{c|c}
K_{\rm star} & {\bf 0} \\
\hline 
{\bf 0} & {\bf 0}
\end{array} 
\right)
\label{SES hamiltonian for half uniform state preparation}
\end{equation}
is a $2N \! \times \! 2N$ block-diagonal Hamiltonian. Here $K_{\rm star}$ is an $N \! \times \! N$ matrix of the form (\ref{K star definition}), and ${\bf 0}$ is the $N \! \times \! N$ zero matrix. The operation time in (\ref{half uniform state preparation identity}) is
\begin{equation}
t_{\rm qc} =\frac{\pi}{\sqrt{N} g_{\rm max}}.
\label{half uniform state preparation time}
\end{equation}
This completes the preparation of the input (\ref{input SES state to adiabatic preparation}) to the adiabatic evolution stage.

At the beginning of the adiabatic evolution stage we program the SES Hamiltonian to be
\begin{equation}
{\cal H} = \left( \begin{array}{c|c}
H_0 & {\bf 0} \\
\hline 
{\bf 0} & {\bf 0}
\end{array} \right),
\end{equation}
where $H_0$ is an $N \! \times \! N$ Hamiltonian with the following properties: 
\begin{enumerate}

\item $H_0$ is real.

\item $[H_0,H]\! \neq \! 0$. 

\item The ground state of $H_0$ is the uniform superposition state (\ref{input SES state to adiabatic preparation}).

\item The ground state is separated from the other eigenstates by an energy gap that is a nondecreasing function of $N$. 

\end{enumerate}
A possible choice when $N$ is a power of two is the ``transverse field" Hamiltonian
\begin{equation}
H_0 = - g_{\rm max} \sum_{i=1}^{\log N} \sigma_i^x.
\label{transverse field}
\end{equation}
However, the explicit matrix forms of (\ref{transverse field}) for large $N$ are complicated and the tensor-product structure is somewhat artifical for our purposes. Instead we use 
\begin{equation}
H_0 = - g_{\rm max} \, K_{\rm full}
\label{preferred initial hamiltonian}
\end{equation}
where $K_{\rm full}$ is an $N \! \times \! N$ matrix of the form (\ref{Kfull definition}). The initial Hamiltonian (\ref{preferred initial hamiltonian}) has eigenvalues
\begin{equation}
E_k = - g_{\rm max} \sum_{j=1}^{N-1} \zeta^{jk}, \ \ \ k\in\lbrace0,1,\cdots,N-1\rbrace,
\end{equation}
where $\zeta \equiv e^{2 \pi i/N}$. The ground state is
\begin{equation}
\frac{1}{\sqrt{N}}
\left( \begin{array}{c}
1  \\
1  \\
\vspace{0.05in} \vdots  \\
1  \\
\end{array} \right) \! ,
\end{equation}
with energy $E_0=-(N-1)g_{\rm max}.$ The remaining eigenfunctions are degenerate with energy $E_{k \neq 0} = g_{\rm max}$.

At later times $0 \le t \le t_{\rm prep}$ the SES Hamiltonian is varied as
\begin{equation}
{\cal H}(t) =\frac{t_{\rm prep}-t}{t_{\rm prep}}  \left( \begin{array}{c|c}
H_0 & {\bf 0} \\
\hline 
{\bf 0} & {\bf 0}
\end{array} \right)
+\frac{t}{t_{\rm prep}} 
\left( \begin{array}{c|c}
\frac{1}{\lambda} H  & {\bf 0} \\
\hline 
{\bf 0} & {\bf 0}
\end{array} \right) \! .
\label{adiabatic Hamiltonian}
\end{equation}
Here
\begin{equation}
\lambda \equiv \frac{  \max_{i i'}  |H_{i i'}|}{g_{\rm max}}
\label{lambda definition for phase estimation}
\end{equation}
is a positive constant that ensures that $\frac{1}{\lambda} H$ can be programmed into the SES processor. This stage of the protocol is standard: In the long $t_{\rm prep}$ adiabatic limit, the processor will be found at $t \! = \! t_{\rm prep}$ in the desired state (\ref{phase estimation initial state}) with high probability.

Next we implement the SES equivalent of the circuit given in Fig.~\ref{IPE circuit figure}, beginning with the Hadamard gate
\begin{equation}
{\sf H} \equiv \frac{1}{\sqrt{2}}
\begin{pmatrix}
1 & 1  \\
1 & -1  \\
\end{pmatrix},
\label{Hadamard gate definition}
\end{equation}
which we write as
\begin{equation}
{\sf H} = - u 
\left( \! \! \begin{array}{cc}
1 & 0 \\
0 & -1 \\
\end{array} \right)
u^\dagger,
\label{Hadamard identity}
\end{equation}
where
\begin{equation}
u \equiv 
\begin{pmatrix}
\sin \frac{\pi}{8} & \cos \frac{\pi}{8} \\
-\!\cos \frac{\pi}{8}  & \sin \frac{\pi}{8}  \\
\end{pmatrix}.
\label{Hadamard u definition}
\end{equation}
Then we have
\begin{equation}
{\sf H} \otimes I_{N \! \times \! N} = 
- (u\otimes I_{N \! \times \! N}) \,  K_z \,
(u^\dagger \otimes I_{N \! \times \! N}),
\end{equation}
where $K_z $ is the $2N \! \times \! 2N$ diagonal matrix
\begin{eqnarray}
K_z &\equiv&
\left( \begin{array}{cccccccc}
 1 & 0& 0&0 & 0& 0& 0& 0\\
\vspace{-0.08in}
0 &  1 & 0&0 & 0& 0& 0& 0\\
0 & 0 &  \ddots &0 & 0& 0& 0& 0\\
0 & 0 & 0& 1  & 0& 0& 0& 0\\
\vspace{-0.08in}
0 & 0 & 0& 0  & -1& 0& 0& 0\\
0 & 0 & 0& 0  & 0& \ddots & 0& 0\\
0 & 0 & 0& 0  & 0& 0& -1& 0\\
0 & 0 & 0& 0  & 0& 0& 0& -1 \\
\end{array} \right),
\label{Kz definition}
\\
&=& - \exp \left[ - i
\left( \begin{array}{cccccc}
\vspace{-0.08in}
\pi & 0  & 0&0 & 0& 0 \\
0 &  \ddots  & 0 &0 & 0&  0\\
0 & 0 & \pi & 0 & 0& 0\\
\vspace{-0.08in}
0 & 0& 0  &  0 & 0& 0\\
0 & 0& 0  & 0& \ddots & 0\\
0 & 0& 0  & 0& 0&  0\\
\end{array} \right) \right].
\label{exponential form of Kz}
\end{eqnarray}
This leads to
\begin{equation}
 e^{-i {\cal H} t_{\rm qc}} = {\sf H} \otimes I_{N \! \times \! N},
\end{equation}
where ${\cal H}$ is the $2N \! \times \! 2N$ Hamiltonian ${\cal H} = g_{\rm max} \, K$, with
\begin{equation}
K \! = \! 
\left( \begin{array}{c|c}
\sin^2(\frac{\pi}{8}) \, I_{N\! \times \! N} & -\cos(\frac{\pi}{8}) \sin(\frac{\pi}{8}) \, I_{N\! \times \! N}   \\
\hline
-\cos(\frac{\pi}{8}) \sin(\frac{\pi}{8}) \, I_{N\! \times \! N} & \cos^2(\frac{\pi}{8}) \, I_{N\! \times \! N}   \\
\end{array} \right)\ \ \ \
\label{SES hamiltonian for Hadamard}
\end{equation}
and
\begin{equation}
t_{\rm qc} = \frac{\pi}{g_{\rm max}}.
\label{SES time for Hadamard}
\end{equation}

The controlled-evolution step has been discussed above in (\ref{general controlled unitary}) through (\ref{controlled exp A identity}). Applying this result to the operation (\ref{controlled evolution operator}) leads to
\begin{equation}
{\rm controlled} \! - \! e^{-iH 2^{m-1} t} 
= \exp \! \left[ -i \left(
\begin{array}{c|c}
{\bf 0} & {\bf 0} \\
\hline 
{\bf 0} & H 
\end{array} 
\right) \! 2^{m-1} t  \right] \! .
\label{SES implementation controlled evolution operator}
\end{equation}
Here ${\bf 0}$ and $H$ are $N \! \times \! N$ matrices, with $H$ the model Hamiltonian, which we assume to be real. Now let $\lambda$ be defined as in (\ref{lambda definition for phase estimation}). Then
\begin{equation}
{\rm controlled} \! - \! e^{-iH 2^{m-1} t}  = e^{-i {\cal H} t_{\rm qc}} ,
\end{equation}
where
\begin{equation}
{\cal H} = \left(
\begin{array}{c|c}
{\bf 0} & {\bf 0} \\
\hline 
{\bf 0} & \frac{1}{\lambda} H
\end{array} 
\right)
\ \ \ {\rm and} \ \ \ 
t_{\rm qc} = \lambda 2^{m-1} t.
\label{SES protocol for controlled evolution operator}
\end{equation}
To perform the controlled-evolution operation, the Hamiltonian in (\ref{SES protocol for controlled evolution operator}) is to be programmed into the SES processor for a time $t_{\rm qc}$.

Finally, we implement the $z$ rotation
\begin{equation}
R_z(\omega)  \otimes I_{N \! \times \! N},
\end{equation}
where
\begin{equation}
R_z(\omega) \equiv e^{-i(\omega/2)\sigma^z}
=
\begin{pmatrix}
e^{-i \omega/2} & 0\\
0 & e^{i \omega/2}   \\
\end{pmatrix}.
\end{equation}
This operation can be generated by applying the $2N \! \times 2N$ Hamiltonian
$ {\cal H} = g_{\rm max} \, K_z $
for a time $t_{\rm qc} = \omega / 2 g_{\rm max}.$

The final stage of the eigenvalue estimation protocol is the SES equivalent of ancilla measurement (see Fig.~\ref{IPE circuit figure}), resulting in the observed value $x_m \in\lbrace 0,1\rbrace$. One way to do this is to perform a simultaneous projective measurement of every qubit in the SES processor. If the excitation at iteration $m$ is observed to be in qubit $i$, we conclude that
\begin{equation}
x_m =
\begin{cases}
0 & \text{if} \ \ 1 \le i \le N, \\
1 & \text{if} \ \ N+1 \le i \le 2N.\\
\end{cases}
\end{equation}
This result follows from the correspondence (\ref{controlled U mapping definition}). The disadvantage of this naive measurement protocol is that it fully collapses the SES wave function, so the eigenfunction $|\psi\rangle$ needs to be re-prepared before the next iteration.

A simple variation of this protocol, however, avoids the state re-preparation step about half the time: Here we simultaneously measure only the first $N \, (=n/2)$ qubits. 
In this case we might observe the excitation to be in qubit $i \in \lbrace 1, 2, \cdots \! , N\rbrace$, or we may not find it at all. Then we conclude that
\begin{equation}
x_m =
\begin{cases}
0 & \text{if the excitation is observed}, \\
1 & \text{if the excitation is not observed.} \\
\end{cases}
\end{equation}
If $x_m\!=\!0$ the measurement fully collapses the state, and we must re-prepare the eigenfunction $|\psi\rangle$. But if $x_m\!=\!1$ we have learned only that the excitation is in the subspace spanned by \begin{equation}
\big\lbrace \big| \textstyle{\frac{n}{2} +1} \big), \big|\textstyle{\frac{n}{2} +2} \big), \cdots \! , \big| n \big) \big\rbrace, 
\end{equation}
which yields no information about $|\psi\rangle$.

It is possible to avoid the eigenfunction re-preparation step altogether by using an example of the {\it ancilla-assisted SES method}: Here we couple an $n$-qubit SES processor to an ancilla qubit, with a degree of connectivity that depends on the application. (The measurement application requires coupling to $n/2$ qubits.) Alternatively, we can regard one of the qubits in a fully connected array of $n+1$ qubits as an ancilla. The essential point is that the device now explores the single-excitation and double-excitation subspaces. The only disadvantage of this measurement protocol is that it requires $N$ steps per iteration.

\begin{figure}
\includegraphics[width=6.0cm]{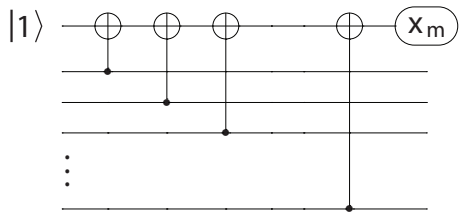} 
\caption{Quantum circuit to measure the parity operator (\ref{parity operator}). The first qubit is an ancilla and the others are the first $N$ qubits of the $2N$-qubit SES processor. The circuit uses $N$ {\sf CNOT} gates.}
\label{parity operator figure}
\end{figure}

The idea is to measure the multi-qubit operator
\begin{equation}
\sigma^z_{1} \otimes \sigma^z_{2} \otimes  \cdots \otimes 
\sigma^z_{N} \otimes I_{N \! \times \! N},
\label{parity operator}
\end{equation}
which projects qubits $i=1,2,\cdots \!,N$ into a state of definite parity. If at iteration $m$ the eigenvalue of (\ref{parity operator}) is observed to be $-1$, the single excitation is in the subspace spanned by $\lbrace |1) , |2), \cdots \! , |N)\rbrace$ and we conclude that $x_m\!=\!0$. If the eigenvalue is $+1$, the excitation is in the space spanned by  $\lbrace |N+1) , |N+2), \cdots \! , |2N)\rbrace$ and we conclude $x_m\!=\!1$. The measurement of the operator (\ref{parity operator}) can be carried out with a single ancilla qubit using the circuit given in Fig.~\ref{parity operator figure}.

It is useful to discuss the nonscalability of the SES method in the context of the eigenvalue estimation application. Typically $N$ is exponentially large in the number of particles, making classical simulation impractical. An {\it ideal} (error-free) quantum computer would require only $O(\log N)$ qubits to run the phase estimation circuit of Fig.~\ref{IPE circuit figure}. However, the large circuit depths required for the controlled evolutions have limited prethreshold applications to very small examples. The SES implementation requires $2N$ qubits, but can perform the controlled evolutions in a single step.

\subsection{Schr\"odinger equation solver for time-independent Hamiltonian matrices}
\label{time-independent Hamiltonian simulator section}

Next we consider the problem of wave function propagation by a real but otherwise arbitrary time-independent Hamiltonian $H$,
\begin{equation}
\big|\psi\big\rangle \rightarrow e^{-i H t} \big|\psi\big\rangle.
\label{time-independent propagation}
\end{equation}
This application, and especially its time-dependent extension discussed below, play to the strengths of the SES chip and suggest a useful prethreshold computational tool. We assume that $H$ is a real, symmetric $n \! \times \! n$ matrix, and we call $H$ the {\it model} Hamiltonian. Here $t$ is the length of simulated time (for example, the duration of some physical process). To map this problem to an SES processor we first find the smallest positive constant $\lambda$ such that every matrix element of
\begin{equation}
{\cal H} = \frac{H - {\rm const} \times I }{\lambda}
\label{rescaled time-independent Hamiltonian construction} 
\end{equation}
is between $-g_{\rm max}$ and $g_{\rm max}$. Here $I$ is the $n \! \times \! n$ identity matrix. When $\lambda > 1$ we are ``compressing" the model Hamiltonian down to that of the SES chip, whereas when $0 < \lambda < 1$ we are expanding it. Such a rescaling is required because the characteristic energy scales of the model and SES chip are usually different. With the SES processor we then perform the equivalent evolution
\begin{equation}
\big|\psi \big\rangle \rightarrow e^{-i {\cal H} t_{\rm qc}} \big| \psi \big\rangle,
\label{scaled evolution}
\end{equation}
where
\begin{equation}
t_{\rm qc} = \lambda \, t.
\label{tqc definition}
\end{equation}
The total time required to perform a single run of the quantum computation is therefore
\begin{equation}
t_{\rm qu}  \equiv t_{\rm qc} +  t_{\rm meas},
\label{total quantum simulation time definition}
\end{equation}
where $t_{\rm meas}$ is the qubit measurement time. For superconducting qubits we can assume $t_{\rm meas}$ to be about $100 \, {\rm ns}$ \cite{SetePRL13}, which includes the time needed for classical post-processing. (Note that the shortest high-fidelity readout time demonstrated to date, including resonator ring-down time, is closer to $300 \, {\rm ns}$ \cite{JeffreyPRL14}. The faster ``catch-disperse-release" protocol of Ref.~\cite{SetePRL13} has not yet been demonstrated.)

A single run of the quantum computer (with readout) simulates a single repetition of an experiment: Initialization, Schr\"odinger evolution, and measurement. It is important to emphasize that such a protocol implements a {\it weak} simulation, providing a single sample from the distribution of possible measurement outcomes, not the probability distributions themselves as is normally computed classically. (This limitation is not specific to the SES method and applies to state propagation with an error-corrected universal quantum computer as well.) For some applications the distinction between weak and strong simulation might be minor. However in other cases it is necessary to estimate the occupation probabilities $p_1, p_2, \dots, p_n$ accurately. We discuss the runtime overhead for strong SES simulation below in Sec.~\ref{strong simulation section}.

How long does a classical simulation of (\ref{time-independent propagation}) take? This of course depends on the model Hamiltonian $H$ (including its dimension $n$ and spectral norm), the value of $t$, and the classical processor and simulation algorithm used. To assess the possibility of quantum speedup, however, it is sufficient to find the minimum time $t_{\rm cl}$ required to classically simulate a given run of an ideal SES processor, with ${\cal H}$ a ``typical" SES Hamiltonian (a real $n \! \times \! n$ random symmetric matrix with all entries between $-g_{\rm max}$ and $g_{\rm max}$), and $t_{\rm qc}$ significantly less than the coherence time. For this analysis we consider the case
\begin{equation}
t_{\rm qc} = 100 \, {\rm ns}
\ \ {\rm and} \ \
\frac{g_{\rm max}}{2\pi} = 50 \, {\rm MHz}.
\end{equation}
The total quantum computation time (\ref{total quantum simulation time definition}) in this example is therefore about
\begin{equation}
t_{\rm qu} = 200 \, {\rm ns}. \ \ \  {\rm (SES \ chip)}
\label{total quantum simulation time}
\end{equation}

We have studied the classical simulation runtime $t_{\rm cl}$ for this problem, comparing, on a single core \cite{computationNote}, three standard numerical algorithms:
\begin{enumerate}

\item {\sl State propagation via Hamiltonian diagonalization.} For a given ${\cal H}$, the unitary matrix $V$ of its eigenvectors and diagonal matrix $D$ of its eigenvalues are first computed. Then we numerically compute the product 
\begin{equation}
V e^{-i D t_{\rm qc}} V^\dagger |\psi\rangle,
\end{equation}
where $|\psi\rangle$ is the initial state. 

\item {\sl Matrix exponentiation via Pad\'e approximation with scaling and squaring} \cite{HighamJMatrixAnalAppl05}. Here we directly compute $\exp(-i {\cal H} t_{\rm qc})$ and then multiply by $|\psi\rangle$. 

\item {\sl Krylov subspace projection} \cite{SidjeTransMathSoft98}. In this case the product $\exp(-i {\cal H} t_{\rm qc}) |\psi\rangle$ itself is directly calculated.

\begin{figure}
\includegraphics[width=8cm]{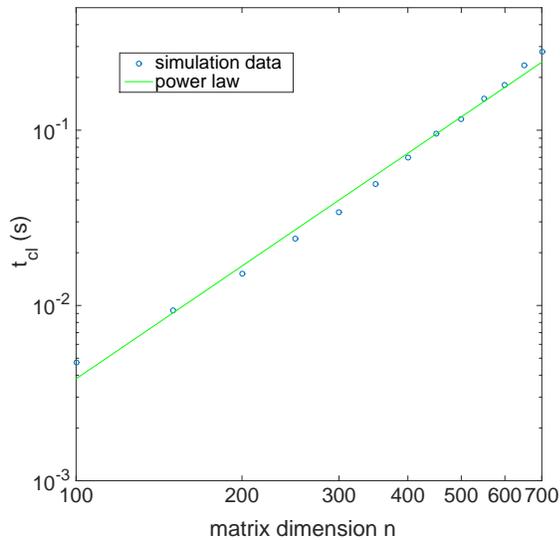} 
\caption{(color online) Classical simulation runtime on a single core \cite{computationNote} versus matrix dimension $n$, in seconds. Here the computational task is solution of the Schr\"odinger equation with a  time-independent Hamiltonian by matrix diagonalization. Data (circles) were determined by averaging the runtimes over 1000 random instances of ${\cal H}$. The solid line is the function $t_{\rm cl}= 203 \! \times \! n^{2.14} \, {\rm ns}$; the scaling becomes $O(n^3)$ at larger $n$. The runtime for a $630 \! \times \! 630$ Hamiltonian is about $ 200 \, {\rm ms}$.}
\label{classical simulation time independent figure}
\end{figure} 

\end{enumerate}
In all cases we assume an initial state $|\psi\rangle$ of the form
\begin{equation}
\left( \begin{array}{c}
1  \\
0  \\
\vspace{0.04in} \vdots  \\
0  \\
\end{array} \right),
\end{equation}
which corresponds to a single SES basis state,
and we average the computation times over 1000 random instances of ${\cal H}$.  Although the three methods have similar speed and accuracy for the particular problem simulated here, the matrix diagonalization method was the fastest, followed by matrix exponentiation. We also tested Runge-Kutta integration and matrix exponentiation via Chebychev polynomial expansion \cite{Tal-EzerJCP84}, which were not competitive with the above methods for the specific application considered. In Fig.~\ref{classical simulation time independent figure} we plot the measured single-core runtimes for the optimal classical algorithm (Hamiltonian diagonalization) versus matrix dimension $n$. We observe that the quantum simulation time (\ref{total quantum simulation time}) is much shorter than all of the single-core runtimes considered.

Our objective is to achieve speedup relative to a state-of-the-art supercomputer, not a single core. The classical simulation runtime $t_{\rm cl}$ should then be evaluated on a supercomputer, using an optimally distributed parallel algorithm. However, we can {\it bound} the parallel performance by using the single-core result and assuming perfect parallelization efficiency: We approximate a petaflop supercomputer by $10^6$ gigaflop cores, and conclude that the classical runtime can be no shorter than $10^{-6}$ times the single-core time. (This is a conservative estimate because high parallelization efficiency is not expected for problem sizes smaller than the number of cores.) We therefore conclude that, for this particular state propagation application, the classical simulation runtime is
no shorter than
\begin{equation}
t_{\rm cl}= 203 \! \times \! n^{2.14} \, {\rm fs}, \ \ \  {\rm (classical \ supercomputer)}
\label{tcl supercomputer time-independent Hamiltonian}
\end{equation}
while the quantum simulation can be performed in a few hundred nanoseconds. The breakeven dimension according to (\ref{tcl supercomputer time-independent Hamiltonian}) is about $n\!=\!630$ qubits. This is quite large given the full connectivity requirement, and it is not known whether such a device could be built in practice. However the breakeven dimension in the time-dependent case (discussed below) is considerably smaller.

In our estimate of the classical simulation runtime we have not included the time needed to store the Hamiltonian matrix in memory or perhaps compute it from a separate procedure. Similarly, for the quantum simulation time estimate we have not included the time required to send the $n(n+1)/2$ controls to the qubits and couplers before the simulation. Furthermore, not every $n > 630$ simulation will exhibit a speedup; this depends on the particular simulated Hamiltonian and the simulated time duration $t$.

An interesting aspect of the Schr\"odinger equation solver is that the complexity is O(1): The quantum simulation time is independent of $n$. This implies that the SES method yields an exponential speedup for this application. However such complexity considerations are probably not meaningful given that the method is not scalable.

\subsection{Schr\"odinger equation solver for time-dependent Hamiltonians: Simulation of molecular collisions}
\label{time-dependent Hamiltonian simulator section}

Finally, we discuss what is perhaps the most interesting application of the SES method known to date, the solution of the Schr\"odinger equation with a time-dependent Hamiltonian. This is a straightforward generalization of the time-independent case, but we expect the time-dependent case to be more useful in practice. In this section we provide a detailed example of time-dependent Hamiltonian simulation with a small SES chip.

\begin{figure}
\includegraphics[width=8cm]{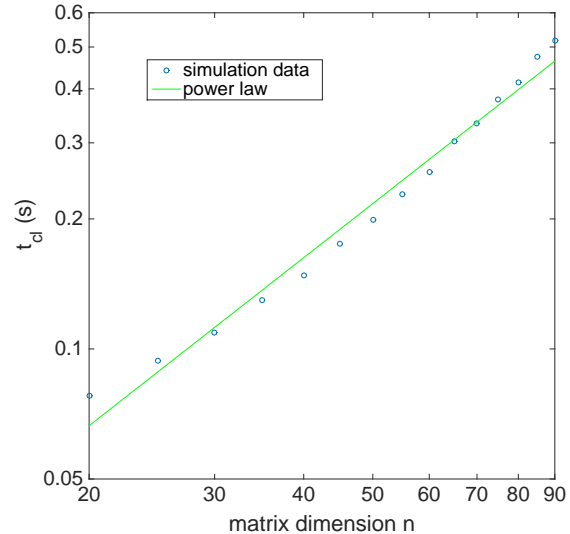} 
\caption{(color online) Classical simulation runtime on a single core \cite{computationNote} versus matrix dimension $n$, in seconds. Here the computational task is solution of the Schr\"odinger equation with a  time-dependent Hamiltonian by Runge-Kutta integration. Data (circles) were determined by averaging the runtimes over 1000 random instances of ${\cal H}$. The solid line is the function $t_{\rm cl}= 1.38 \times n^{1.29} \, {\rm ms}$. The simulation time for a $50 \! \times \! 50$ Hamiltonian is about $ 200 \, {\rm ms}$.}
\label{classical simulation time dependent figure}
\end{figure}

Time-dependent Hamiltonian simulation is implemented by varying the SES matrix elements (\ref{SES hamiltonian}) according to some protocol, which can be done with nanosecond resolution. This does not require any additional runtime, the time complexity is still constant, and the total quantum simulation runtime for a $100 \, {\rm ns}$ evolution is again given by (\ref{total quantum simulation time definition}). Although the classical runtime is problem specific, we can again assess the possibility of speedup by estimating the time required to classically simulate an ideal SES processor, in this case with all $n^2$ matrix elements varying on a nanosecond timescale. There are two types of numerical simulation algorithms we consider: 
\begin{enumerate}

\item {\sl Runge-Kutta integration.} Here we solve the system of coupled ordinary differential equations
\begin{equation}
{\dot a} = -i {\cal H} a.
\end{equation}
Although the Runge-Kutta runtime is slower than diagonalization for a time-independent Hamiltonian, it does not slow down significantly when ${\cal H}$ is time dependent.

\item {\sl Time slicing combined with diagonalization.} This algorithm is based on an approximate decomposition of the time-dependent problem into a sequence of constant-Hamiltonian  intervals, each of width $\Delta t$. The time $\Delta t$ must be significantly smaller than the characteristic timescale of matrix element variation, for example  $\Delta t = 0.1 \, {\rm ns}$. Then 
\begin{equation}
N_{\rm slice} = \frac{t_{\rm qc}}{\Delta t}
\end{equation}
time slices are required, and the classical runtime using this approach will be approximately $N_{\rm slice}$ times longer than (\ref{tcl supercomputer time-independent Hamiltonian}). For the $100 \, {\rm ns}$ evolution, $N_{\rm slice}  \! = \! 1000$.
\end{enumerate}
We find that Runge-Kutta integration is the fastest approach for the specific problem considered here. In Fig.~\ref{classical simulation time dependent figure} we plot the measured single-core runtimes for the optimal classical algorithm (Runge-Kutta integration) versus matrix dimension $n$. Bounding the performance of this algorithm on a petaflop supercomputer by including a factor of $10^{-6}$ (recall discussion from Sec.~\ref{time-independent Hamiltonian simulator section}), we conclude that for this application the classical simulation runtime is no shorter than
\begin{equation}
t_{\rm cl}= 1.38 \! \times \! n^{1.29} \, {\rm ns}, \ \ \  {\rm (classical \ supercomputer)}
\label{tcl supercomputer time-dependent Hamiltonian}
\end{equation}
while the quantum simulation can be performed in a few hundred nanoseconds. The breakeven dimension according to (\ref{tcl supercomputer time-dependent Hamiltonian}) is around $n\!=\!50$ qubits. As expected, this value is much smaller than the breakeven in the time-independent case. {\it An SES Schr\"odinger equation solver of modest size might be able to achieve quantum speedup relative to a petaflop supercomputer.}

We turn now to a detailed example of time-dependent Hamiltonian simulation with a small programmable SES chip. One particularly interesting application of the method is to the quantum simulation of atomic and molecular collisions. Collisions are especially well suited for SES simulation because they typically involve modest Hilbert spaces---tens to thousands of channels---and in the time-dependent formulation involve Hamiltonians that are naturally bounded in time. In particular, the initial and final asymptotic Hamiltonians for neutral scatterers are diagonal (in the adiabatic basis), whereas the off-diagonal elements rapidly turn on and then off during the collision itself, inducing transitions between the channels. Although the Born-Oppenheimer potential energy surfaces used here do require a classically inefficient electronic structure precomputation, the largest potential energy surface calculations \cite{BraamsIRPC09,BowmanPCCP11} are far ahead of the largest classical collision simulations performed to date \cite{WangJCP06,SchiffelJCP10,YangNatComm15}. SES implementation of the semiclassical Born-Oppenheimer problem therefore has the potential to push molecular collision simulations to new unexplored regimes. (We note that there are related chemical reaction simulation methods developed by Lidar {\it et al.}~\cite{LidarPRE99} and by Kassal {\it et al.}~\cite{KassalPNAS08} that do not require a precomputed potential surface, but these require an error-corrected quantum computer to implement and are not prethreshold methods.) Another useful feature of the scattering application is the convenient mapping of each molecular channel to a single SES basis state, which is possible because of the similar way initial states are prepared and final states measured in both the processor and a collision experiment. We also find that the atomic physics time and energy scales turn out to map nicely to that of superconducting qubits after optimal rescaling.

\begin{figure}
\includegraphics[width=8cm]{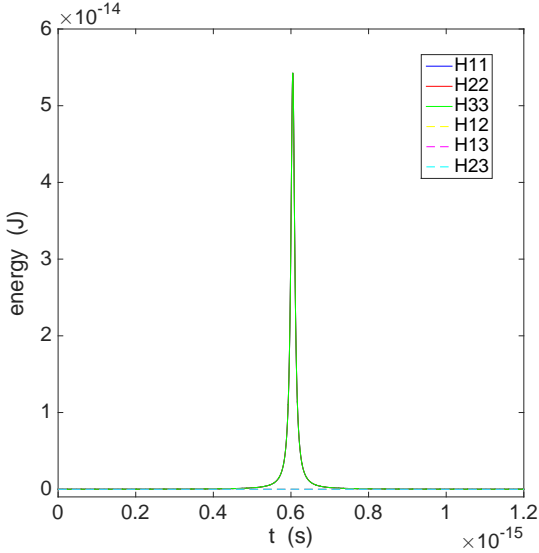} 
\caption{(color online) Time dependence of the matrix elements of the scattering Hamiltonian for a collision with $v_0\!=\!2.0$ and $b\!=\!0.5$ in atomic units. The diagonal matrix elements (solid curves) are similar in magnitude and cannot be resolved in this figure. The off-diagonal elements (dashed curves) are much smaller than the diagonal elements and also cannot be resolved here. In this example the collision energy is $E_{\rm cm} \! = \! 341 \, {\rm keV}$ and the impact occurs at $t_0 \! = 6 \! \times \! 10^{-16} \, {\rm s}$.}
\label{scattering Hamiltonian vs time figure}
\end{figure} 

To illustrate this application we consider a three-channel Na-He collision (an unpublished preliminary account of this application is given in Ref.~\cite{PritchettEtalArxiv1008.0701}). The three channels included in our model and their correspondence with SES basis states are
\begin{equation}
{\rm Na(3s) + He(1s^2)} \  [1 \ ^{2}\Sigma^{+}]
\ \ \longleftrightarrow \ \ 
 \big|1\big) 
\label{ground channel}
\end{equation}
and
\begin{equation}
{\rm Na(3p) + He(1s^2)} \  [1 \ ^{2}\Pi; 2 \ ^{2}\Sigma^{+}]
\ \ \longleftrightarrow \ \ 
\big|2\big), \big|3\big).
\label{excited channels}
\end{equation}
The square brackets indicate the molecular structure of the channels. In this model, the helium atom remains in its electronic ground state $1{\rm s}^2$ during the collision (the excitation energies of its excited states are too high to be relevant here), whereas sodium can be excited from its ground state $3{\rm s}$ to either of two excited states, both denoted by $3{\rm p}$. In the physical system, the channels (\ref{ground channel}) and (\ref{excited channels}) have additional degeneracies, including spin degeneracies, but they do not affect the collision probabilities calculated here. Precomputed Born-Oppenheimer energies and nonadiabatic couplings of the Na-He system \cite{LinPRA08} are stored for fixed values of the internuclear distance $R$, and we make a standard semiclassical (high energy) approximation and assume that the scatterers follow a straight-line trajectory. Then, for an impact occuring at a time $t_0$, the internuclear separation varies according to
\begin{equation}
R = \sqrt{b^2 + v_0^2 (t-t_0)^2},
\label{straight line trajectory}
\end{equation}
where $v_0$ is the initial relative velocity and $b$ the impact parameter of the collision. The relative velocity is related to the collision energy in the center-of-mass frame through $E_{\rm cm} = \mu v_0^2 / 2,$ where $\mu$ is the reduced mass. The procedure outlined in Appendix~\ref{scattering Hamiltonian section} then leads to the scattering Hamiltonian shown in Fig.~\ref{scattering Hamiltonian vs time figure}.

\begin{figure}
\includegraphics[width=8cm]{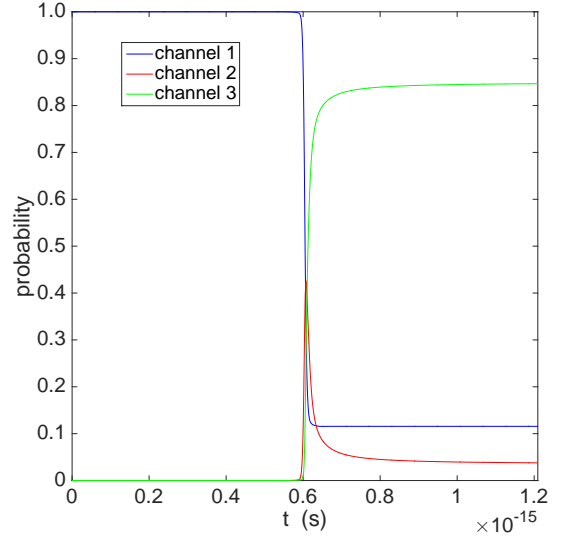} 
\caption{(color) Scattering probabilities $p_{1 \rightarrow i}(t)$ for a Na-He collision with $v_0\!=\!2.0$ and $b\!=\!0.5$ in atomic units. The system is initially prepared in channel 1. The collision occurs at $t_0 \! = \! 6 \! \times \! 10^{-16} \, {\rm s}$.}
\label{scattering probabilities figure}
\end{figure} 

In Fig.~\ref{scattering probabilities figure} we plot the probabilities $p_{1 \rightarrow i}(t)$ for the Na-He system to be found in channel $i \in \lbrace 1,2,3 \rbrace$ after being initially prepared in channel 1, the ground state. The final values $p_{1 \rightarrow i}(\infty)$ are the probabilities for an elastic ($i \! = \! 1$) or inelastic ($i \! = \! 2,3$) collision with a given $E_{\rm cm}$ and $b$, which we find to be
\begin{eqnarray}
p_{1 \rightarrow 1} &=& 0.116 \nonumber \\
p_{1 \rightarrow 2} &=& 0.038 \nonumber \\
p_{1 \rightarrow 3} &=& 0.846.
\label{final scattering probabilities}
\end{eqnarray}

To simulate this process with a programmable SES chip we must first rescale the physical or model Hamiltonian so that it fits on the SES processor. Doing this optimally is critical to the utility of any time-dependent simulation so we will discuss it in some detail: Suppose for the moment that our model Hamiltonian is given by a time-independent $n \! \times \! n$ real symmetric matrix $H$. Dividing $H$ by any positive constant $\lambda$ while rescaling the evolution time by the same factor obviously leaves the dynamics invariant. Because we want the quantum simulation to be as fast as possible, we choose the smallest value of $\lambda$ that makes $H/\lambda$ compatible with the SES processor (every matrix element of $H/\lambda$ is between $-g_{\rm max}$ and $g_{\rm max}$). As mentioned above, if $\lambda>1$ we are compressing the model's energy scales to fit on the SES chip, whereas if $\lambda<1$ we are expanding them. This naive approach to rescaling, however, does not take advantage of the fact that we can always shift $H$ by a constant (which changes the corresponding states by a phase factor that we do not measure). Including this gauge transformation results in the rescaling used above in (\ref{rescaled time-independent Hamiltonian construction}). The time $t_{\rm qc}$ required to simulate an evolution of duration $t$ is simply given by (\ref{tqc definition}), but now we will go further and regard (\ref{tqc definition}) as giving the linear relationship between the simulated and physical times {\it during} a process. To generalize this construction to a time-dependent model Hamiltonian $H(t),$ we first compute the mean of its diagonal elements, 
\begin{equation}
c(t) \equiv \frac{1}{n}\sum_{i=1}^n H_{ii},
\label{mean of diag elements}
\end{equation}
and then find, at each time $t,$ the smallest positive $\lambda$ such that every matrix element of
\begin{equation}
{\cal H}(t) = \frac{H(t) - c(t) \times I }{\lambda(t)}
\label{rescaled time-dependent Hamiltonian construction} 
\end{equation}
is between $-g_{\rm max}$ and $g_{\rm max}$. The function $\lambda(t)$ defines the resulting {\it nonlinear} relation between the physical and simulated times according to
\begin{equation}
t_{\rm qc}(t) = \int_0^t \lambda \ dt'.
\label{nonlinear scaling definition}
\end{equation}
Equation (\ref{rescaled time-dependent Hamiltonian construction}) gives the simulated Hamiltonian as a function of the physical time $t$, and (\ref{nonlinear scaling definition}) is then inverted to find the desired ${\cal H}(t_{\rm qc})$, which in turn is programmed into the SES chip.

\begin{figure}
\includegraphics[width=8cm]{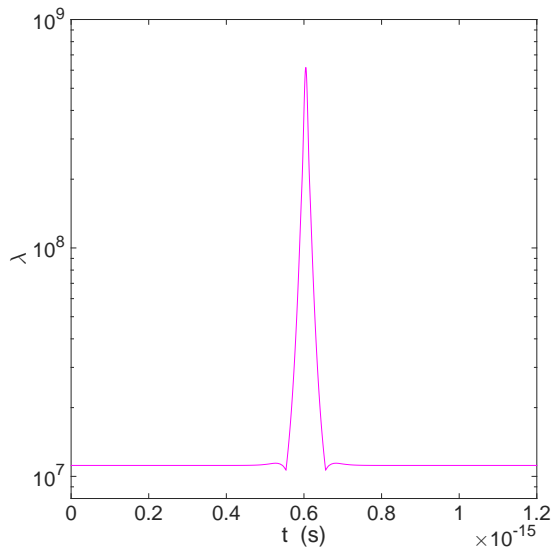} 
\caption{(color online) Rescaling function for the Na-He collision simulation with $g_{\rm max}/2\pi \! = \! 30 \, {\rm MHz}$. Collision parameters are $v_0\!=\!2.0$ and $b\!=\!0.5$ in atomic units. We find that $\lambda$ has an asymptotic value around $10^7$ and reaches $6 \! \times \! 10^8$ during the collision.}
\label{lambda figure}
\end{figure} 

\begin{figure}
\includegraphics[width=8cm]{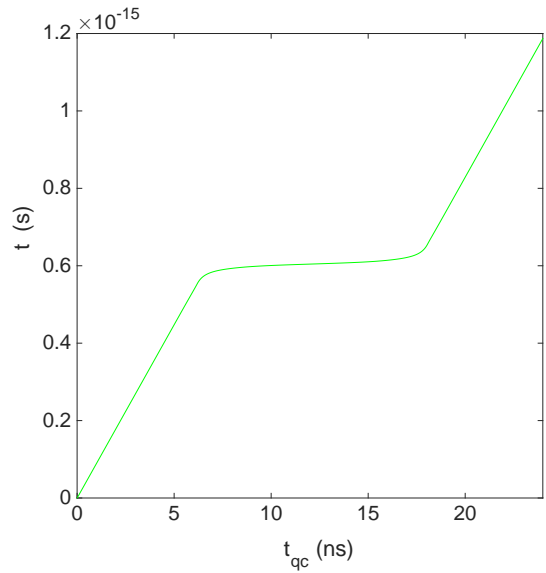} 
\caption{(color online) Nonlinear time scaling for the $\lambda(t)$ given in Fig.~\ref{lambda figure}. Most of the simulation time is spent near the moment of collision, $t_0$, and a single run of the simulation is completed in $24 \, {\rm ns}$.}
\label{time vs tQC figure}
\end{figure} 

Applying this procedure to the Na-He collision problem results in the rescaling function $\lambda$ shown in Fig.~\ref{lambda figure}, and the nonlinear time relationship shown in Fig.~\ref{time vs tQC figure}. We note that the nonlinear energy/time rescaling protocol is extremely effective at mapping this atomic physics problem to the SES processor, allowing a single run of the simulation (excluding measurement) to be completed in about $24 \, {\rm ns}$. The $\lambda$ function shown in Fig.~\ref{lambda figure} assumes $g_{\rm max}/2\pi \! = \! 30 \, {\rm MHz}$; if this is increased to $50 \, {\rm MHz}$ the simulation is completed in $15 \, {\rm ns}$. It is important to emphasize that any positive piecewise continuous function $\lambda(t)$ defines a mathematically valid energy/time rescaling, and that the specific form used in practice should be determined by hardware considerations, such as qubit coherence times and control pulse bandwidth. In particular, $\lambda(t)$ can be chosen to bound both the magnitude of the SES matrix elements and their rates of change, but we will not pursue this variation here.

\begin{figure}
\includegraphics[width=8cm]{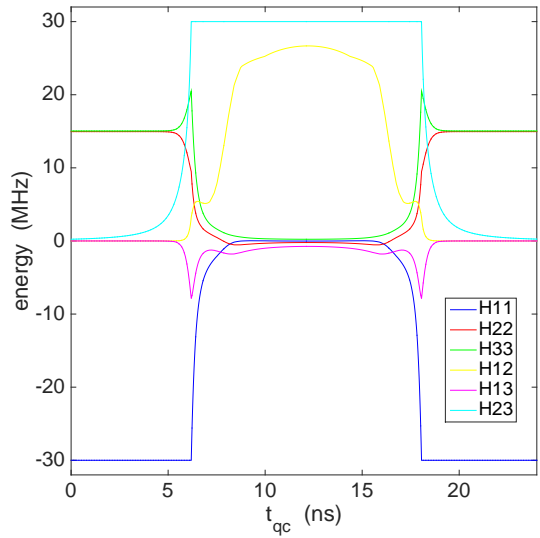} 
\caption{(color) SES Hamiltonian matrix elements for a Na-He collision with $v_0\!=\!2.0$ and $b\!=\!0.5$, in atomic units. Here we assume that $g_{\rm max}/2\pi \! = \! 30 \, {\rm MHz}$. At each instant the magnitude of at least one matrix element achieves its maximum value of $30 \, {\rm MHz}$, making the simulation as fast as possible.}
\label{scattering Hamiltonian vs time on QC figure}
\end{figure} 

\begin{figure}
\includegraphics[width=8cm]{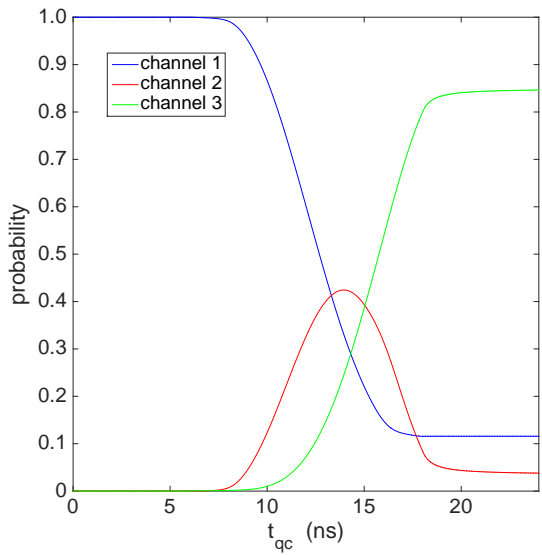} 
\caption{(color) Scattering probabilities $p_{1 \rightarrow i}(t)$ in the SES processor for a Na-He collision with $v_0\!=\!2.0$ and $b\!=\!0.5$ in atomic units. The dynamics near the moment of collision occupy most of the simulation.}
\label{scattering probabilities on QC figure}
\end{figure}

Use of the rescaling function given in Fig.~\ref{lambda figure} leads to the SES matrix elements shown in Fig.~\ref{scattering Hamiltonian vs time on QC figure}, which bear no resemblance to those of Fig.~\ref{scattering Hamiltonian vs time figure}. The corresponding scattering probabilities during the simulation are shown in Fig.~\ref{scattering probabilities on QC figure}. Compared with Fig.~\ref{scattering probabilities figure}, we see that the dynamics near the moment of collision are rescaled to occupy most of the simulation. The final scattering probabilities are the same as in Fig.~\ref{scattering probabilities figure} and are given in (\ref{final scattering probabilities}).

\subsection{Strong quantum simulation}
\label{strong simulation section}

As we have emphasized, a single run of the SES chip provides a single sample from the distribution of possible measurement outcomes---a weak simulation---not the probability distributions themselves as is normally computed classically. If the objective is to perform a strong simulation and hence estimate the $n$ basis state occupation probabilities $p_i$, it is necessary to repeat the simulation and readout many times. In this section we discuss the runtime overhead for strong SES simulation. (The objective considered is that of measuring occupation probabilities, not probability amplitudes.)

Suppose that after some simulation we want to measure the occupation probability $p$ of one qubit in the SES chip. We do this by performing the simulation $N$ times, after each repetition $r$ measuring the qubit in the diagonal basis and observing $x \in \lbrace 0,1 \rbrace$. The estimate
\begin{equation}
p_{\rm est} = \frac{1}{N} \sum_{r=1}^N x_r 
\label{estimated value of p}
\end{equation}
for $p$ will have a sampling error (standard error of the mean) given by
\begin{equation}
{\sf E} \equiv \sqrt{{\rm var}\big(p_{\rm est}\big)} = \sqrt{\frac{p(1-p)}{N}} \le \frac{1}{2 \sqrt{N}}.
\label{single qubit sampling error}
\end {equation}
For example, to ensure that the sampling error is smaller than 1\%, it is sufficient to repeat the simulation 2500 times. 

The overhead $N \! = \! 2500$ is a worst-case estimate for a 1\% sampling error. If $p$ is known to be small (or close to 1), fewer repetitions are required. However in this work we restrict ourselves to the case where measuring $p$ is the objective of a quantum simulation and is not known a priori.

We turn now to the runtime overhead for estimating every SES basis state occupation probability $p_i$ in an $n$-qubit processor. The qubits are measured simultaneously and the sampling error formula (\ref{single qubit sampling error}) apples to each $p_i$. The only modification resulting from the SES constraint---that the device is in the single-excitation subspace---is that the $n$ probabilities are not independent, because
\begin{equation}
\sum_{i=1}^n p_i = 1.
\label{excitation probability sum rule}
\end{equation} 
However the condition (\ref{excitation probability sum rule}) does not affect the sampling statistics. Therefore we conclude that the sampling error for the $i^{\rm th}$ qubit is
\begin{equation}
{\sf E}_i = \sqrt{\frac{p_i(1-p_i)}{N}}.
\label{qubit i sampling error}
\end {equation}

The result (\ref{qubit i sampling error}) shows that the strong simulation overhead, given by the required number of repetitions $N$, is independent of $n$ and is no worse than that for a single qubit. In particular, the upper bound  ${\sf E}_i \le (2\sqrt{N})^{-1}$ applies, and hence the required number of repetitions satisfies
\begin{equation}
N \le  \frac{1}{4 {\sf E}_i^2 }.
\label{repetitions bound}
\end {equation}
In conclusion, the complexity for strong quantum simulation is also {\it constant}, and the overhead factor $N$ is no larger than the single-qubit value. 

Although the sampling error can be made arbitrarily small, it is usually not helpful to require it to be smaller than the deviations in the occupation probabilities resulting from other error sources, such as decoherence. Consider, for example, the evolution of a single qubit for 100ns in the presence of $T_1$ relaxation. If $T_1 \! = \! 40{\rm \mu s}$, decoherence would lead to a 0.25\% error in the excited state probability $p$. It is not possible, in general, to measure the original qubit excitation probability with better than 99.75\% accuracy because of this error. Thus, decoherence limits the accuracy of a strong simulation independently of the sampling error. We will discuss decoherence and other error sources in detail in the next section.

\section{ACCURACY OF SES COMPUTATION}
\label{error section}

In this section we discuss the errors incurred during a quantum computation or simulation with an $n$-qubit SES chip. We are especially interested in the $n$-dependence of these errors, for large $n$, and whether they pose any serious limitation to the practical utility of the SES approach (we conclude that they do not). Below we separately analyze decoherence errors, matrix element ``control" errors, and leakage out of the SES. In each case the ideal, error-free state after some process is an SES pure state $| \psi_{\rm ideal}  \rangle$, and we estimate the error
\begin{equation}
{\sf E} \equiv 1 -  \langle \psi_{\rm ideal} | \rho | \psi_{\rm ideal}  \rangle ,
\label{SES error definition}
\end{equation}
where $\rho$ is the realized final state. 

\subsection{Energy relaxation error}

The first decoherence error we discuss is energy relaxation (zero-temperature amplitude damping). We estimate this error by setting ${\cal H}_{i i'} = 0$ and  calculating the decay of an initially prepared SES state
\begin{equation}
| \psi_{\rm init} \rangle = \sum_{i=1}^n a_i \, | i ) 
= \sum_{i=1}^n a_i \, | 0 \cdots 1_i \cdots 0 \rangle
\label{initial SES state}
\end{equation}
in the absence of unitary evolution.

The single-qubit Kraus matrices for this process are
\begin{equation}
E_1 =
\begin{pmatrix}
1 & 0 \\
0 & \sqrt{r} \\
\end{pmatrix}
\ \ {\rm and} \ \
E_2 =
\begin{pmatrix}
0 & \sqrt{1-r} \\
0 & 0 \\
\end{pmatrix},
\label{Kraus matrices energy relaxation}
\end{equation}
with
\begin{equation}
r = e^{-t_{\rm qc}/T_1}.
\end{equation}
Here $t_{\rm qc}$ is the runtime for some SES quantum computation and all qubits are assumed to have the same $T_1$ value. The $T_1$ time for capacitively coupled Xmon qubits is currently about $40 \, {\rm \mu s}$ \cite{BarendsPRL13}.

In the presence of energy relaxation,
\begin{equation}
\rho_{\rm init} \equiv | \psi_{\rm init} \rangle \langle \psi_{\rm init} |
\rightarrow E_1^{\otimes n} \rho_{\rm init} \ 
E_1^{\dagger \otimes n} + \cdots, 
\end{equation}
where the dots denote terms involving one or more applications of the $E_2$ operator (\ref{Kraus matrices energy relaxation}) that are outside of the SES and do not contribute to (\ref{SES error definition}). Then
\begin{equation}
\rho_{\rm init} \rightarrow r \, \rho_{\rm init} + \cdots
\end{equation}
and
\begin{equation}
{\sf E} = 1 - e^{-t_{\rm qc}/T_1} \approx \frac{t_{\rm qc}}{T_1},
\label{SES error energy relaxation}
\end{equation}
which is independent of $n$. An SES state (in the absence of unitary evolution) thefore relaxes at the same rate as a single excited qubit. The approximation in (\ref{SES error energy relaxation}) applies when $t_{\rm qc} \ll T_1$, which is the regime of interest here.

\subsection{Pure dephasing error}

Next we discuss pure dephasing, which in Xmon qubits is believed to be caused primarily by flux noise (however this has been recently questioned \cite{OMalleyPhysRevApp15}). We again estimate this error by setting ${\cal H}_{i i'} = 0$ and calculating the degradation of an initially prepared SES state (\ref{initial SES state}) in the absence of unitary evolution. We assume a standard single-qubit dephasing model with no correlations between the noise at different qubits.

The Kraus matrices in this case are
\begin{equation}
E_1 =
\begin{pmatrix}
1 & 0 \\
0 & \sqrt{r} \\
\end{pmatrix}
\ \ {\rm and} \ \
E_2 =
\begin{pmatrix}
0 & 0 \\
0 & \sqrt{1-r} \\
\end{pmatrix},
\label{Kraus matrices pure dephasing}
\end{equation}
with
\begin{equation}
r = e^{-2t_{\rm qc}/T_\phi}.
\end{equation}
We can estimate the $T_\phi$ time for Xmon qubits (with fixed capacitive coupling) by using the relation
\begin{equation}
\frac{1}{T_\phi} = \frac{1}{T_2} - \frac{1}{2 T_1},
\end{equation}
with values $T_1 = 40 \, {\rm \mu s}$ and $T_2 = 20 \, {\rm \mu s}$ from Ref.~\cite{BarendsPRL13}, which leads to $T_\phi \approx 27 \, {\rm \mu s}$. (Note, however, that the flux noise in Xmon qubits is not Markovian, as we have assumed. We believe that our simple dephasing calculation {\it overestimates} the actual dephasing error.) All qubits are assumed to have the same $T_\phi$ value.

In the presence of pure dephasing,
\begin{eqnarray}
&&\rho_{\rm init} \rightarrow E_1^{\otimes n} \rho_{\rm init} \ 
E_1^{\dagger \otimes n} \nonumber  \\
&+& \sum_{i=1}^n \, (E_1 \! \otimes \cdots \! \!  \!\underbrace{E_2}_{{\rm qubit} \, i} \!  \!  \! \cdots \otimes \!  E_1) \, \rho_{\rm init}  \,
(E_1 \! \otimes \cdots E_2 \cdots \otimes \!  E_1)^\dagger ,
\nonumber
\end{eqnarray}
where the $E_{1,2}$ now refer to (\ref{Kraus matrices pure dephasing}), and we have used the fact that terms with two or more applications of the $E_2$ operator vanish when applied to $\rho_{\rm init}$. Note that $E_2$ annihilates an SES basis state $|i) = |0\cdots 1_i \cdots 0\rangle$ unless it acts on qubit $i$, in which case it produces a factor of $\sqrt{1-r}$. Then dephasing transforms an SES state to
\begin{equation}
\rho_{\rm init} \rightarrow r \, \rho_{\rm init} 
+ (1-r) \sum_{i=1}^n |a_i|^2 \, |i)(i|,
\end{equation}
and the associated fidelity loss is
\begin{eqnarray}
{\sf E} &=& \big(1 - e^{-2 t_{\rm qc} /T_\phi} \big) \bigg( 1 - \sum_{i=1}^n |a_i|^4 \bigg) \nonumber \\
&\approx& \frac{2 t_{\rm qc}}{T_\phi} \bigg( 1 - \sum_{i=1}^n |a_i|^4 \bigg).
\label{SES error pure dephasing}
\end{eqnarray}

The dephasing error (\ref{SES error pure dephasing}) is maximized when the SES basis states are equally populated, $|a_i | = 1/\sqrt{n}$. Then
\begin{equation}
\max_{a_i} \, {\sf E} = (1 - e^{-2t_{\rm qc}/T_\phi}) \bigg( 1 - \frac{1}{n} \bigg)
\approx \frac{2 t_{\rm qc}}{T_\phi} \bigg( 1 - \frac{1}{n} \bigg).
\label{max dephasing error}
\end{equation}
This expression is valid for $n \ge 1$ (the $n \! = \! 1$ SES state $|1\rangle$ has no pure dephasing error). The $n$-dependence of the worst-case dephasing error (\ref{max dephasing error}) is very favorable, approaching a constant as $n \rightarrow \infty$. Therefore the pure dephasing error is bounded by
\begin{equation}
{\sf E}  \le  1 - e^{-2 t_{\rm qc} /T_\phi} \approx \frac{2 t_{\rm qc}}{T_\phi},\label{dephasing error bound}
\end{equation}
which is only a few times larger than (\ref{SES error energy relaxation}).

The total error due to decoherence is the sum of (\ref{SES error energy relaxation}) and (\ref{SES error pure dephasing}) [or (\ref{dephasing error bound})]. Assuming a $t_{\rm qc} \! =\!100 \, {\rm ns}$ computation and the coherence times given above, this error is around $1\%$ and is independent of $n$. 

\subsection{Hamiltonian control errors}

Next we calculate the error (\ref{SES error definition}) caused by imperfect experimental programming of the SES matrix elements, which we call a control error. We assume that the intended Hamiltonian is a real, symmetric,  time-independent $n \! \times \! n$  matrix ${\cal H}$, but that the applied Hamiltonian is instead ${\cal H} + V$, where $V$ is a real, symmetric, time-independent matrix that does not commute with ${\cal H}$.

We consider a typical situation where the processor is initially prepared in a single SES basis state $|i)$. The nonideal final state (neglecting decoherence) is then
\begin{equation}
e^{-i ({\cal H}+V) t } | i ),
\end{equation}
where $t$ is the evolution time for either a complete algorithm or a single step in an algorithm. The error (\ref{SES error definition}) in this case is therefore
\begin{equation}
{\sf E}_i = 1 - \big| (i| e^{i {\cal H} t} e^{-i ({\cal H}+V) t} | i ) \big|^2 \! .
\label{state i control error}
\end{equation} 
Averaging (\ref{state i control error}) over the initial SES basis state leads to
\begin{equation}
{\sf E} = 1 - \frac{1}{n}\sum_{i=1}^n \big| (i| e^{i {\cal H} t} e^{-i ({\cal H}+V) t} | i ) \big|^2.
\label{control error definition}
\end{equation} 
The SES Hamiltonian ${\cal H}$ in (\ref{control error definition}) is assumed to have the ``typical" form described above in Sec.~\ref{SES Hamiltonian section}.

In this section we evaluate the control error (\ref{control error definition}) using two complementary approaches. First we consider the small-$Vt$ perturbative limit. Evaluating quantities of the form
\begin{equation} 
U \equiv e^{i {\cal H} t} e^{-i ({\cal H}+V) t}
\label{U definition}
\end{equation}
by a series expansion in $V$ is a standard problem in perturbation theory: Differentiating (\ref{U definition}) with respect to time yields
\begin{equation}
i \frac{\partial U}{\partial t} = {\tilde V}(t) \, U, 
\ \ \ {\rm with} \ \ \ 
{\tilde V}(t) \equiv e^{i H t} V e^{-i H t},
\end{equation}
showing that (\ref{U definition}) satisfies a Schr\"odinger equation with time-dependent  Hamiltonian ${\tilde V}$. We can therefore write (\ref{U definition}) as
\begin{equation} 
U = T \, e^{- i \int_0^t {\tilde V}(t') \, dt'} ,
\label{time-ordered exponential}
\end{equation}
where $T$ is the time-ordering operator. Expanding (\ref{time-ordered exponential}) to second order yields
\begin{widetext}
\begin{eqnarray}
U &=& 1 - i \! \int_0^t dt_1 \, {\tilde V}(t_1) 
- \frac{1}{2} \! \int_0^t \! dt_1 \! \! \int_0^t \! dt_2 \,
T \big( {\tilde V}(t_1)  {\tilde V}(t_2) \big) \nonumber \\
&=& 1 - i \int_0^t \! dt_1 \, {\tilde V}(t_1) 
- \frac{1}{2} \! \int_0^t \! dt_1
\bigg( \! \int_0^{t_1} \! \! dt_2 \,  {\tilde V}(t_1)  {\tilde V}(t_2)
 + \! \int_{t_1}^t \! dt_2 \,  {\tilde V}(t_2)  {\tilde V}(t_1) \bigg). 
\end{eqnarray}
This leads to
\begin{eqnarray} 
( i | U | i ) &=& 
1 - i \int_0^t \! dt_1 \, ( i | e^{i {\cal H} t_1} V e^{-i {\cal H} t_1} |i ) \, 
 - \frac{1}{2} \! \int_0^t \! dt_1
\bigg[ \! \int_0^{t_1} \! \! dt_2 \,  ( i | e^{i {\cal H} t_1} V 
e^{-i {\cal H} (t_1-t_2)} V e^{-i {\cal H} t_2} | i )
\nonumber \\
 &+& \int_{t_1}^t \! dt_2 \,  ( i | e^{i {\cal H} t_2} V e^{i {\cal H} (t_1 - t_2)} V e^{-i {\cal H} t_1} | i )  \bigg]
=
1 - i \sum_{jk} \int_0^t \! dt_1 \, S_{ij}(-t_1) \, S_{ki}(t_1) \, V_{jk}
\nonumber \\
 &-& \frac{1}{2} \sum_{jklm} \int_0^t \! dt_1
\bigg[ \! \int_0^{t_1} \! \! dt_2 \,  
S_{ij}(-t_1) \, S_{kl}(t_1-t_2) \, S_{mi}(t_2)
 + \int_{t_1}^t \! dt_2 \,  
 S_{ij}(-t_2) \, S_{kl}(t_2-t_1) \, S_{mi}(t_1)  \bigg]
 V_{jk} V_{lm} ,
\end{eqnarray}
where
\begin{equation}
S_{i i'}(t) \equiv ( i | e^{-i {\cal H} t} | i' ).
\label{S definition}
\end{equation}
Because ${\cal H}$ is symmetric, $S_{i i'}(t)$ is also a symmetric matrix. Then to second order in $V$ we have
\begin{eqnarray} 
\big| ( i | U | i ) \big|^2 &=& 1 + 2 \, {\rm Im} \sum_{jk}
\int_0^t dt_1 \, S_{ij}(-t_1) \, S_{ki}(t_1) \, V_{jk}
+ \! \sum_{j j' k k'} \int_0^t dt_1 \! \int_0^t dt'_1 \,
S_{ij}(-t_1) \, S_{ki}(t_1) \, S_{ij'}^*(-t'_1) \, S^*_{k' i}(t'_1)
\, V_{jk} \, V_{j' k'} 
\nonumber \\
&-& {\rm Re} \, \sum_{j k l m} \int_0^t dt_1 \bigg[
\int_0^{t_1} \! dt_2 \,
S_{ij}(-t_1) \, S_{kl}(t_1 - t_2) \,  S_{mi}(t_2)
+ \int_{t_1}^{t} \! dt_2 \,
S_{ij}(-t_2) \, S_{kl}(t_2 - t_1) \,  S_{mi}(t_1)
\bigg] V_{jk} \, V_{lm}. \ \ \ \ \ \ \ \
\label{second order propagator}
\end{eqnarray}
The control error to second order for a fixed ${\cal H}$ and $V$ follows from (\ref{control error definition}) and (\ref{second order propagator}).

\end{widetext}

Next we regard the perturbation as random and average over the random matrix $V$. The elements $V_{i \le i'}$ are assumed to be independent identically distributed random variables (the elements $V_{i>i'}$ fixed by symmetry). The moments of $V$ are
\begin{equation}
\overline{\, V_{ab}} = 0 
\label{random matrix first moment}
\end{equation}
and
\begin{equation}
\overline{\, V_{ab}V_{cd}} = \sigma^2 \big( \delta_{ac} \delta_{bd} + \delta_{ad} \delta_{bc}  -  \delta_{ac} \delta_{bd}  \delta_{ab} \big),
\label{random matrix second moment}
\end{equation}
where $\sigma$ is a parameter with dimensions of energy characterizing the size of the control errors. The condition (\ref{random matrix second moment}) enforces the symmetry requirement of $V$. The perturbation-averaged control error (\ref{control error definition}) is then found to be
\begin{eqnarray}
{\sf E} &=& \sigma^2 t^2 + \frac{\sigma^2}{n} \! \int_0^t \! dt_1 \, dt_2 
\bigg[ \big| {\rm Tr} \, S(t_1 - t_2) \big|^2
\nonumber \\
&-& 2 \sum_{i=1}^n \big| S_{ii}(t_1 - t_2) \big|^2 
- \sum_{i=1}^n \big| S_{ii}(t_1 + t_2) \big|^2
\nonumber \\
&+& \sum_{i,j=1}^n \big| S_{ij}(t_1) \big|^2 \, \big| S_{ij}(t_2) \big|^2
\bigg],
\label{state-averaged perturbation-averaged error}
\end{eqnarray}
where the propagator $S_{i i'}$ is defined in (\ref{S definition}). The control error is proportional to $\sigma^2$, as expected, and for fixed evolution time $t$ and large $n$ is dominated by the second term (the first term in the square brackets). By retaining this  dominant second term, together with the first $\sigma^2 t^2$ term, and evaluating the trace in the eigenfunction basis, we can perform the time integrations analytically to obtain a useful spectral form for the perturbative control error,
\begin{equation}
{\sf E} \approx 2 \sigma^2 t^2 + \frac{2\sigma^2}{n} \! \!\sum_{\begin{scriptsize}
\begin{array}{c}
\alpha, \alpha^\prime \\
\alpha \! \neq \!  \alpha^\prime
\end{array}
\end{scriptsize}}
\frac{1-\cos[(E_{\alpha}-E_{\alpha^\prime})t]}{(E_{\alpha}-E_{\alpha^\prime})^2},
\label{spectral form of perturbative control error}
\end{equation}
where the $E_\alpha$ are the eigenvalues of ${\cal H}$. 

The expression ({\ref{spectral form of perturbative control error}) is useful for studying the $n$-dependence of the control error for short times, corresponding to a single step in an SES computation. An example is shown in  
Fig.~\ref{control error spectral form figure} for $t \! = \! 10 \, {\rm ns}$ and matrix elements of the perturbation $V$ uniformly distributed in the interval
\begin{equation}
\bigg( \! \! - \! \frac{\delta V}{2} , \frac{\delta V}{2} \bigg),
\end{equation}
with $\delta V / 2 \pi = 0.5 \, {\rm MHz}$. The value of $\sigma$ for this distribution is 
\begin{equation}
\sigma = \frac{\delta V}{\sqrt{12} \, }.
\end{equation}
We conclude from Fig.~\ref{control error spectral form figure} that the expression ({\ref{spectral form of perturbative control error}) provides an accurate approximation for the short-time control errors, and that they are less than 0.1\% in all the cases considered.

\begin{figure}
\includegraphics[width=8.0cm]{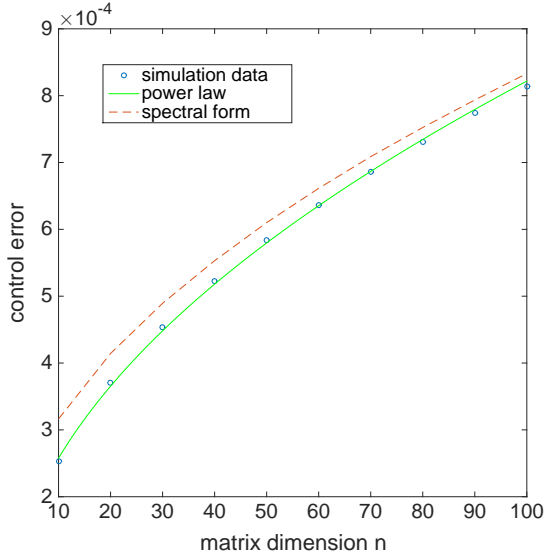} 
\caption{(color online) Plot of the control error {\sf E} versus matrix dimension for $g_{\rm max} / 2 \pi \!=\!50 \, {\rm MHz}$, $t\!=\!10 \, {\rm ns},$ and errors in the individual matrix elements uniformly distributed between -0.25 and $0.25 \, {\rm MHz}$. Data (open circles) follow from an exact numerical calculation of (\ref{control error definition}), averaged over ideal Hamiltonian ${\cal H}$ and perturbation $V.$ The solid curve is the function ${\sf E} = 8.1 \! \times \! 10^{-5} \! \times \! n^{0.50}$. The dashed curve follows from the expression (\ref{spectral form of perturbative control error}).}
\label{control error spectral form figure}
\end{figure} 

The perturbative result (\ref{state-averaged perturbation-averaged error}) can also be used to understand the long-time asymptotic limit of the control error. For fixed dimension $n$ and large evolution time $t$, we find that
\begin{equation}
{\sf E} \approx 2 \sigma^2 t^2,
\label{long time limit of control error}
\end{equation}
independent of $n$. Both the first $\sigma^2 t^2$ term in (\ref{state-averaged perturbation-averaged error}) and the second term contribute to (\ref{long time limit of control error}). From (\ref{long time limit of control error}) we obtain an upper bound 
\begin{equation}
t_{\rm max} = \frac{1}{\sqrt{2} \sigma}
\label{perturbation theory time limit}
\end{equation}
on the evolution time $t$ for which perturbation theory is valid.

It is also useful to directly calculate the control error (\ref{control error definition}) numerically, which is useful for exploring the nonperturbative regime. The results are shown in Figs.~\ref{control error 100 figure} and \ref{control error 1000 figure}. The most important conclusions of these simulations is that the dimension dependence of the control error grows much more slowly than linearly, and for errors in the SES matrix elements $\le  \! 0.25 \, {\rm MHz}$ is about 2\% when $n\!=\!100$ and 3\% when $n\!=\!1000.$ If the size of the errors in the matrix elements are doubled---to $\le  \! 0.50 \, {\rm MHz}$---the $n\!=\!100$ control error increases to 7.4\%, less than that predicted by the quadratic $\sigma^2$ scaling resulting from perturbation theory.

\begin{figure}
\includegraphics[width=8.0cm]{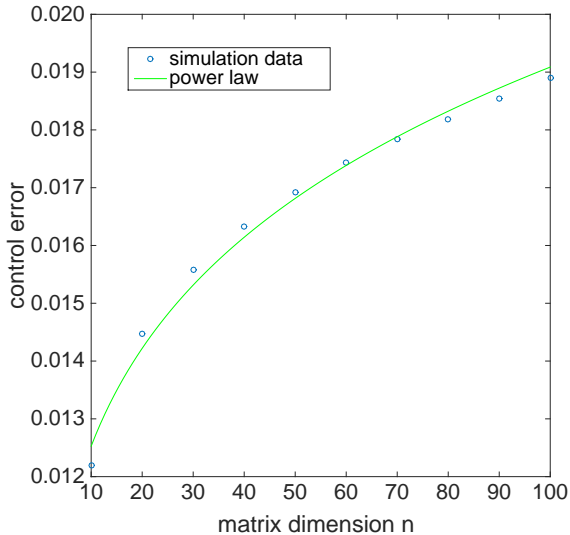} 
\caption{(color online) Fidelity loss caused by control errors in the SES Hamiltonian versus number of qubits or matrix dimension $n$. Here we assume $g_{\rm max} / 2 \pi \!=\!50 \, {\rm MHz}$, $t\!=\!100 \, {\rm ns},$ and errors in the individual matrix elements uniformly distributed between -0.25 and $0.25 \, {\rm MHz}$. Errors (open circles) are averaged over ideal Hamiltonian ${\cal H}$ and perturbation $V.$ The solid line is the function ${\sf E} = 8.2 \! \times \! 10^{-3} \! \times \! n^{0.18}$.}
\label{control error 100 figure}
\end{figure} 

\begin{figure}
\includegraphics[width=8.0cm]{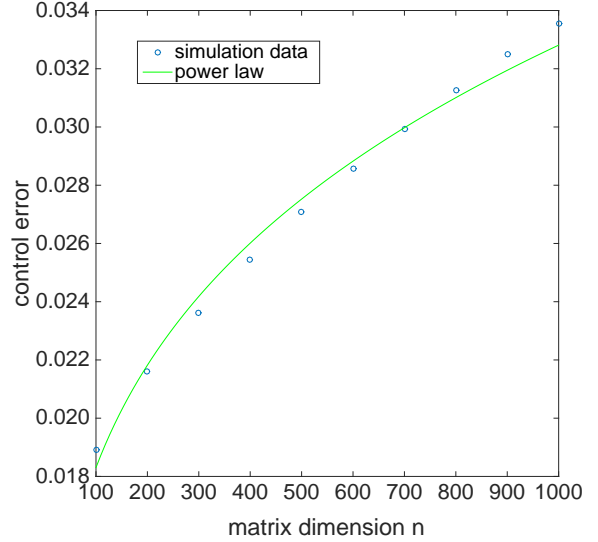} 
\caption{(color online) Same as Fig.~\ref{control error 100 figure} for larger SES matrix dimension. Here $g_{\rm max} / 2 \pi \!=\!50 \, {\rm MHz}$, $t\!=\!100 \, {\rm ns},$ and errors in the individual matrix elements uniformly distributed between -0.25 and $0.25 \, {\rm MHz}$. The solid line is the function ${\sf E} = 5.7 \! \times \! 10^{-3} \! \times \! n^{0.25}$.}
\label{control error 1000 figure}
\end{figure} 

\subsection{Leakage out of the SES}

We briefly comment on two additional error mechanisms that result in excitation out of the SES, namely leakage into the triple-excitation subspace and $|2\rangle$-state errors. To understand the origin of leakage into the triple-excitation subspace, we use the identity [see (\ref{c definition})] $\sigma^x = c + c^\dagger$ and expand the qubit-qubit interaction in (\ref{QC model}). Terms proportional to $c_i^\dagger c_{i'}$ and $c_i c^\dagger_{i'}$ connect SES states to other SES states, whereas the terms proportional to $c_i^\dagger c_{i'}^\dagger$ connect SES states to triply excited states. However these excitations are protected by large energy gaps and the corresponding errors are negligible. 

To understand the origin of $|2\rangle$-state errors, recall that
with conventional gate-based superconducting quantum computation, the dominant sources of $|2\rangle$-state excitation are microwave pulses and two-qubit gates. However neither of these are used in the SES approach. Initialization of SES basis states can produce $|2\rangle$-state errors, but it is known how to limit these errors to less than $10^{-4}$ \cite{MotzoiPRL09}.

\section{CONCLUDING REMARKS}
\label{conclusion section}

The SES method described here appears to be distinct from previously investigated approaches to quantum computation. Like analog quantum simulation, it is tied to a specific hardware model and cannot be implemented on any architecture. However, the approach enables universal quantum computation and simulation, and might make  quantum speedup possible with prethreshold hardware.

To understand the origin of the speedup, we introduce a fictional quantum computer model consisting of $q$ qubits and a Hamiltonian containing ($i$ times) every element of the Lie algebra ${\rm su}(2^q)$, with independent experimental control over each of its $2^{2q}-1$ elements. Let's call this a  {\it supercharged} quantum computer. A supercharged quantum computer is capable of implementing any operation in ${\rm SU}(2^q)$ by a single application of the Hamiltonian, bypassing the need to decompose such operations into elementary one- and two-qubit gates. It is clear that a supercharged quantum computer is more powerful than a traditional  universal quantum computer: It can efficiently perform any computation that is in BQP (defined with respect to a traditional quantum computer), but it can also {\it efficiently} perform some quantum computations that are outside of BQP. In particular, a unitary chosen randomly from ${\rm SU}(2^q)$ has no polynomial-depth gate decomposition, but can be implemented by a supercharged quantum computer in constant time. An SES processor with $n\!=\!2^q$ qubits and capable of implementing arbitrary {\it complex} Hamiltonians would be able to  simulate a supercharged quantum computer. The programmable SES processor introduced here, which can only implement real Hamiltonians, is somewhat less powerful than a supercharged quantum computer, but for many applications complex Hamiltonians are not required and the supercharged quantum computer model correctly explains why quantum speedup is possible with the SES method.

It is interesting to notice how decoherence only barely limits the problem sizes that can be implemented with a programmable SES processor, at least for the applications explored here. The main factor limiting the utility of the SES method is the difficulty of building fully connected arrays of qubits. In this sense we can say that the SES approach trades the familiar limitations resulting from decoherence for a new limitation---that of building qubit graphs with high connectivity. Relative to the large, community-wide effort devoted to studying and improving quantum coherence, the problem of increasing connectivity is certainly in its infancy.

If we accept that the SES method outperforms the traditional gate-based approach for prethreshold universal quantum computation, but that it is ultimately unscalable, the question becomes whether an SES chip could be built that is large enough to be of practical use (and before an error-corrected universal quantum computer arrives). We speculate that for selected applications, breakeven with a classical million-core supercomputer is possible, a significant feat, but that a stronger form of speedup---such as performing computations that are essentially impossible classically---is probably not. Perhaps a breakeven-sized SES processor would be useful for its low power consumption or as a special-purpose subunit in a conventional error-corrected quantum computer.

\begin{acknowledgments}

This work was supported by the National Science Foundation under CDI grant DMR-1029764. It is a pleasure to thank 
Chris Adami,
Alan Aspuru-Guzik, 
Yu Chen,
Robert Geller,
Joydip Ghosh, 
Nadav Katz,
Ronnie Kosloff,
Peter Love,
Matteo Mariantoni,
Anthony Megrant,
Charles Neill,
Pedram Roushan,
Anton Zeilinger,
and
Zhongyuan Zhou
for useful discussions.
Part of this work was carried out while M.G.~was a Lady Davis Visiting Professor in the Racah Institute of Physics at Hebrew University, Jerusalem.

\end{acknowledgments}

\appendix

\section{General qubit-qubit coupling types}
\label{general coupling types section}

In this section we discuss the generalization of the SES method to fully connected quantum computer  models of the form
\begin{eqnarray}
H_{\rm qc} = \sum_{i} \epsilon_i c_i^\dagger c_i + \frac{1}{2} \sum_{i i'} 
g_{ii'} \sum_{\mu \nu}  \, J_{\mu \nu} \, \sigma^\mu_i\otimes\sigma^\nu_{i'}, \ \ \ 
\label{general interaction model}
\end{eqnarray}
where the $\sigma^\mu$ (with $\mu=x,y,z$) are Pauli matrices and $J_{\mu \nu}$
is a fixed, real, dimensionless tensor. In this case the SES matrix elements are
\begin{eqnarray}
\big( i \big| H_{\rm qc} \big|i' \big) &=& \bigg[ \epsilon_i  
- 2 \big( \sum_{j} g_{ij} \big) J_{zz} 
+  \big(\sum_{j<j'} g_{jj'}\big) J_{zz}  \bigg] \delta_{ii'} \nonumber \\
&+& \bigg[J_{xx} + J_{yy} - i (J_{xy}-J_{yx}) \bigg] g_{ii'}.
\label{general matrix elements}
\end{eqnarray}
Note that the term proportional to $(\sum_{j<j'} g_{jj'})\delta_{ii'}$ is an energy 
shift and can be dropped. The SES method can be applied (possibly with some protocol modifications) whenever 
\begin{equation}
J_{xx} + J_{yy}  \neq 0 
\label{exchange condition}
\end{equation}
and 
\begin{equation}
J_{xy}=J_{yx} \, .
\label{J symmetry condition}
\end{equation}
The condition (\ref{exchange condition}) means that the interaction has an exchange or transverse component, and (\ref{J symmetry condition}) ensures that the SES Hamiltonian is purely real.

\section{Tunable coupler circuit}
\label{tunable coupler circuit section}

\begin{figure}
\includegraphics[width=8.0cm]{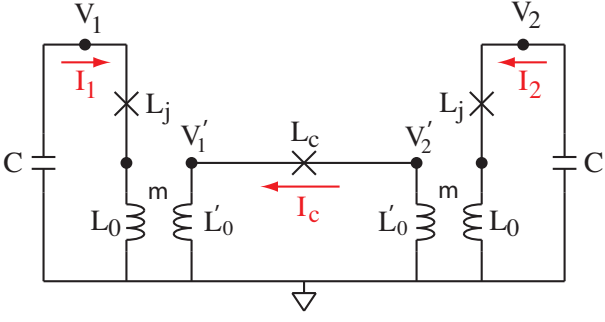} 
\caption{(Color online) Basic coupler circuit for SES processor. The three crosses represent Josephson junctions, each with a flux-tunable Josephson coupling energy. Each Xmon qubit has capacitance $C$ and tunable junction inductance $L_{\rm j}$. The coupler wire has a Josephson junction with tunable inductance $L_{\rm c}$ and mutual inductance $m$ as indicated. Self inductances of the coils are denoted by $L_{0}$ and $L'_{0}$.}
\label{SES coupler 2 qubits figure}
\end{figure} 

In this section we calculate the qubit-qubit interaction strength $g$ for the coupler circuit shown in Fig.~\ref{SES coupler 2 qubits figure}, which is the building block for a programmable SES chip. We first give a simplified treatment by making weak coupling and harmonic approximations, and then discuss the general case afterward.

The circuit of Fig.~\ref{SES coupler 2 qubits figure} has six active nodes (black dots) and is described by six node flux coordinates \cite{devoretLesHouches97}. However, all nodes except those labeled $V_{1,2}$ have negligible capacitance to ground and are therefore ``massless" degrees of freedom that remain in their instantaneous ground states. They will be eliminated from the problem in the analysis below. The Lagrangian for the circuit of Fig.~\ref{SES coupler 2 qubits figure} is
\begin{equation}
L= \sum_{i=1,2}  \bigg(\frac{\Phi_0}{2\pi}\bigg)^{\! \! 2}  
\frac{C}{2} {\dot \varphi_i}^2 \ - \ U,
\label{circuit lagrangian}
\end{equation}
where $\Phi_0 \equiv h/2e$ is the flux quantum, $C$ is the qubit capacitance, $\varphi_{1,2}$ is the dimensionless node flux at $V_{1,2}$, and $U$ is the total potential energy. Following the approach of Ref.~[\onlinecite{ChenEtalPRL14}], we replace the inductive network of Fig.~\ref{SES coupler 2 qubits figure} (excluding the capacitors) by the equivalent circuit of Fig.~\ref{SES coupler black box figure}, where $M$ and $L_{\rm q}$ are effective inductances to be determined in terms of the physical circuit parameters.

\begin{figure}
\includegraphics[width=7.0cm]{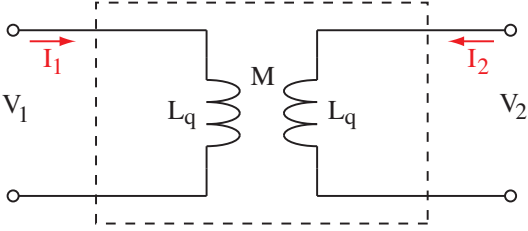} 
\caption{(Color online) Replacing the inductive network of Fig.~\ref{SES coupler 2 qubits figure} by an effective circuit with mutual inductance $M$ and self inductances $L_{\rm q}$.}
\label{SES coupler black box figure}
\end{figure} 

The potential energy in the circuit of Fig.~\ref{SES coupler black box figure} is $L_{\rm q}( I_1^2 + I_2^2)/2 + M I_1 I_2.$ Rewriting this in terms of magnetic flux
\begin{equation}
\begin{pmatrix}
\Phi_1 \\
\Phi_2 \\
\end{pmatrix}
= 
\begin{pmatrix}
L_{\rm q} & M \\
M & L_{\rm q} \\
\end{pmatrix}
\begin{pmatrix}
I_1 \\
I_2 \\
\end{pmatrix}
\label{flux current relation}
\end{equation}
we have
\begin{equation}
U = \frac{\Phi_1^2}{2K L_{\rm q}} + \frac{\Phi_2^2}{2K L_{\rm q}} + \Gamma_{11} \, \Phi_1 \Phi_2,
\label{magentic energy expression}
\end{equation}
where
\begin{equation}
K = 1- \bigg(\frac{M}{L_{\rm q}} \bigg)^2 
\ \ \ {\rm and} \ \ \
\Gamma_{11} = - \frac{M}{K L_{\rm q}^2}.
\label{K and Gamma11 definitions}
\end{equation}
The cross term in (\ref{magentic energy expression}) proportional to $\Gamma_{11}$ is responsible for the qubit-qubit coupling. In the weak coupling limit, $M \ll L_{\rm q}$ and $K \approx 1$, which we assume below.

Next we calculate the qubit-qubit interaction strength $g$ induced by this cross term. Perhaps the simplest way to do this is to use the expression 
\begin{equation}
\Phi = \frac{1}{\sqrt{2 \epsilon C}} \, (a + a^\dagger)
\label{quantized flux operator}
\end{equation}
for the flux of an LC oscillator with frequency
\begin{equation}
\epsilon  = \frac{1}{\sqrt{L_{\rm q} C}},
\label{epsilon definition} 
\end{equation}
in terms of creation and annihilation operators. This leads to the desired result \cite{ChenEtalPRL14}
\begin{equation}
g =  \frac{\Gamma_{11} L_{\rm q}}{2} \, \epsilon.
\label{g in terms of Gamma11}
\end{equation} 

Finally, we find $M$ and $L_{\rm q}$ in terms of the physical circuit parameters. Assuming an $e^{i \omega t}$ time dependence we have from 
(\ref{flux current relation}) that
\begin{equation}
M = \frac{1}{i \omega} \! \times \! \bigg( \frac{V_1}{I_2} \bigg)_{\! \! I_1 = 0} 
\ \ \ {\rm and} \ \ \
L_{\rm q} = \frac{1}{i \omega} \! \times \! \bigg( \frac{V_1}{I_1} \bigg)_{\! \! I_2 = 0} \! \! .
\label{M and Lq definition}
\end{equation}
Using these expressions we find
\begin{equation}
M = \frac{m^2}{L_{\rm c} + 2 L_0^\prime}
\ \ \ {\rm and} \ \ \
L_{\rm q} = L_{\rm j} + L_0 - M,
\label{M and Lq in terms of circuit parameters}
\end{equation}
from which we obtain
\begin{equation}
g = -\frac{m^2}{2(L_{\rm j} + L_0)(L_{\rm c} + 2 L_0^\prime)} \, \epsilon.
\label{coupling strength result}
\end{equation}
Here $\epsilon$ is the qubit frequency (\ref{epsilon definition}). This expression for the strength of the transverse $\sigma^x \otimes \sigma^x$ coupling in the weak coupling and harmonic approximations is the main result of this section. The advantage of the circuit of Fig.~\ref{SES coupler 2 qubits figure} over that of Ref.~[\onlinecite{ChenEtalPRL14}] when extended to many qubits is the absence of coupler loops through which the flux must be individually controlled.

Based on our recent work [\onlinecite{GellerDonateChenEtalPre14}] on a closely related coupler circuit, we expect the result (\ref{coupling strength result}) to be a good approximation to the actual coupling. The main difference is that the qubit anharmonicity suppresses the magnitude of the coupling (in Ref.~[\onlinecite{GellerDonateChenEtalPre14}] the coupling was found to be suppressed by about 15\%). Anharmonicity also generates a small ($\lesssim \! 1 \, {\rm MHz}$) diagonal $\sigma^z \otimes \sigma^z$ interaction, but such an interaction has no effect on a single excitation.

\section{Scattering Hamiltonian}
\label{scattering Hamiltonian section}

\begin{figure}
\includegraphics[width=8cm]{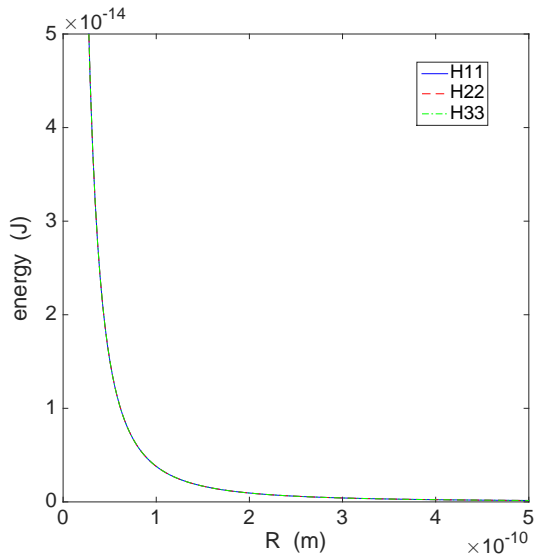} 
\caption{(color online) Diagonal elements $H_{ii}$ of the Na-He scattering Hamiltonian with collision parameters $v_0\!=\!2.0$ a.u. and $b\!=\!0.5$ a.u. The three curves cannot be resolved in this figure.}
\label{scattering Hamiltonian diag vs R figure}
\end{figure}

In this section we outline the construction of the scattering Hamiltonian for the Na-He collision discussed in Sec.~\ref{time-dependent Hamiltonian simulator section}. We begin by constructing a $3 \! \times \! 3$ matrix $U$ as a function of internuclear distance $R$, which we refer to as the potential-coupling matrix, and which is written in atomic units (a.u.). The atomic unit of energy is the Hartree $E_{\rm h} \, $ ($\approx 4.36 \! \times \! 10^{-18} \, {\rm J}$), the atomic unit of length is the Bohr radius $a_0$ ($\approx 5.29 \! \times \! 10^{-11} \, {\rm m})$, and the atomic unit of time is $\hbar / E_{\rm h}$ ($\approx   2.42 \! \times \! 10^{-17} \, {\rm s}$). The diagonal elements of $U$ are the three diabatic potential energies shown in Fig.~1 of Lin {\it et al.} \cite{LinPRA08}, using the molecular state basis given in (\ref{ground channel}) and (\ref{excited channels}), converted to atomic units. The element $U_{13}$ is the diabatic radial coupling given in Fig.~3a of Ref.~\cite{LinPRA08}. The element $U_{12}$ is the diabatic rotational coupling shown as a dashed line in Fig.~3b of \cite{LinPRA08} (and incorrectly labeled there as $2' \ ^{2}\Sigma^{+} - 1' \ ^{2}\Pi$), and  $U_{23}$ is the rotational coupling shown as a solid line in Fig.~3b (and incorrectly labeled $1' \ ^{2}\Sigma^{+} - 1' \ ^{2}\Pi$). 

\begin{figure}
\includegraphics[width=8cm]{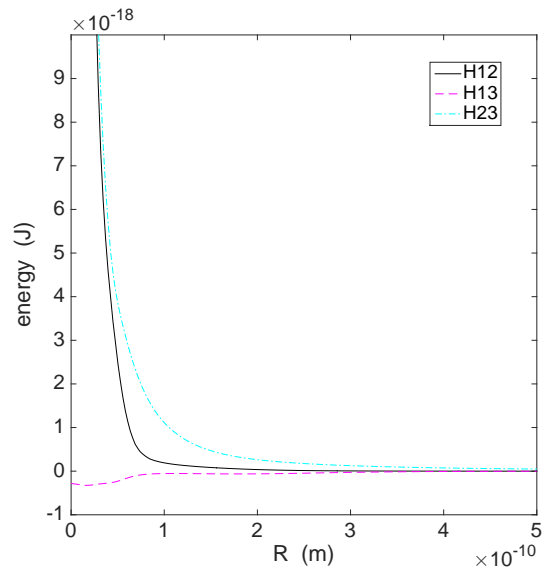} 
\caption{(color online) Off-diagaonal elments $H_{i i'}$ of the Na-He scattering Hamiltonian with collision parameters $v_0\!=\!2.0$ a.u.~and $b\!=\!0.5$ a.u.}
\label{scattering Hamiltonian offdiag vs R figure}
\end{figure} 

The potential-coupling matrix is used to define the {\it scattering Hamiltonian} $H$, also expressed in atomic units, as follows: The diagonal elements are given by
\begin{equation}
H_{ii}(R) = U_{ii}(R) +  \frac{\mu}{2} \bigg( \frac{b v_0}{R} \bigg)^2 \! \! , \ \ \ \ \ i=1,2,3
\label{diagonal scattering Hamiltonian definition}
\end{equation} 
where $U_{ii}(R)$ are the diabatic electronic potentials discussed above, $\mu \! = \! 6214.35 \, {\rm a. u.}$ is the Na-He reduced mass, $b$ is the impact parameter, and $v_0$ is the initial relative velocity of the collision (all in atomic units). The additional centrifugal terms in (\ref{diagonal scattering Hamiltonian definition}) are obtained by making the classical approximation that the orbital angular momentum $\hbar\ell$ is equal to $b\mu v_0,$ with $\ell \gg 1$ and $\ell(\ell +1)\approx \ell^2$. Due to the high kinetic energy considered here, the collision dynamics and scattering probabilities are not affected by the centrifugal terms, which only produce an energy shift. The diagonal elements of $H$ in SI units are plotted in Fig.~\ref{scattering Hamiltonian diag vs R figure}. 

The off-diagonal rotational coupling elements are given by
\begin{eqnarray}
H_{12}(R) &=&  \bigg( \frac{b v_0}{R^2} \bigg) \times U_{12}(R), \label{H12 definintion} \\
H_{23}(R) &=&  \bigg( \frac{b v_0}{R^2} \bigg) \times U_{23}(R), 
\label{H23 definintion}
\end{eqnarray} 
where the classical approximation to the orbital angular momentum is again applied, and the off-diagonal radial coupling element is given by
\begin{equation}
H_{13}(R) = U_{13}(R).
\label{H13 definintion}
\end{equation} 
These are plotted in SI units in Fig.~\ref{scattering Hamiltonian offdiag vs R figure}. Our definitions of $U$ and $H$ follow from Eqs.~(4-35) and (4-47b) of Ref.~\cite{BransdenChargeExchange}.

The $R$-dependent scattering Hamiltonian (\ref{diagonal scattering Hamiltonian definition}) through (\ref{H13 definintion}) becomes time-dependent after assuming the semiclassical trajectory (\ref{straight line trajectory}). Scatterers are assumed to have initial and final internuclear separations of $R \! = \! 50 \, {\rm a.u.}$, resulting in the time-dependent Hamiltonian shown in Fig.~\ref{scattering Hamiltonian vs time figure}.

\bibliography{../../bibliographies/algorithms,../../bibliographies/applications,../../bibliographies/dwave,../../bibliographies/control,../../bibliographies/error_correction,../../bibliographies/general,../../bibliographies/group,../../bibliographies/ions,../../bibliographies/math,../../bibliographies/nmr,../../bibliographies/optics,../../bibliographies/simulation,../../bibliographies/superconductors,../../bibliographies/surface_code,endnotes}

\begin{thebibliography}{87}
\bibitem [{opt()}]{opticalSesNote}%
  \BibitemOpen
  \href@noop {} {}\bibinfo {note} {As discussed, we have in mind a Hamiltonian
  specified in some basis---a real symmetric matrix with no particular
  structure. Cases where $e^{-iHt}$ has a known decomposition into a short
  sequence of elementary operations can also be simulated optically using the
  methods of Refs.~\cite{ReckPRL94} and \cite{CerfPRA98}.}\BibitemShut {Stop}%
\bibitem [{res()}]{resonatorNote}%
  \BibitemOpen
  \href@noop {} {}\bibinfo {note} {The state preparation requirement could be
  achieved by coupling one of the resonators to an auxiliary qubit, or by using
  a complete graph of $n-1$ resonators together with a single qubit. Dispersive
  readout techniques can be applied to resonators as well as
  qubits.}\BibitemShut {Stop}%
\end{thebibliography}

\end{document}